\documentclass[aps,prb]{revtex4}
\usepackage{psfig} 
\input{macros.phys}

\def\sinf{\sigma_{\infty}}

\def\eh{{\bf \hat{e}}}
\def\fg{{\bf f}}

\begin{document}

\title{Random tilings of high symmetry: I. Mean-field theory}

\author{N. Destainville}
\affiliation{Laboratoire de Physique Th\'eorique, UMR CNRS-UPS 5152,
Universit\'e Paul Sabatier,
31062 Toulouse Cedex 04, France.}

\author{M. Widom}
\affiliation{Department of Physics, Carnegie Mellon University, 
Pittsburgh, PA 15213, USA.}

\author{R. Mosseri} \affiliation{Laboratoire de Physique Th\'eorique des
Liquides, Tour 24, Bo\^{\i}te 121, 4, Place Jussieu, 75252 Paris Cedex 05,
France.}

\author{F. Bailly}
\affiliation{LPSB-CNRS, 92195 Meudon Cedex, France.}

\date{\today}

\begin{abstract}
We study random tiling models in the limit of high rotational
symmetry. In this limit a mean-field theory yields reasonable
predictions for the configurational entropy of free boundary rhombus
tilings in two dimensions. We base our mean-field theory on an
iterative tiling construction inspired by the work of de Bruijn. In
addition to the entropy, we consider correlation functions, phason
elasticity and the thermodynamic limit. Tilings of dimension other
than two are considered briefly.
\end{abstract}

\maketitle

\section{Introduction}
\label{intro}

\begin{figure}[b]
\begin{center}
\ \psfig{figure=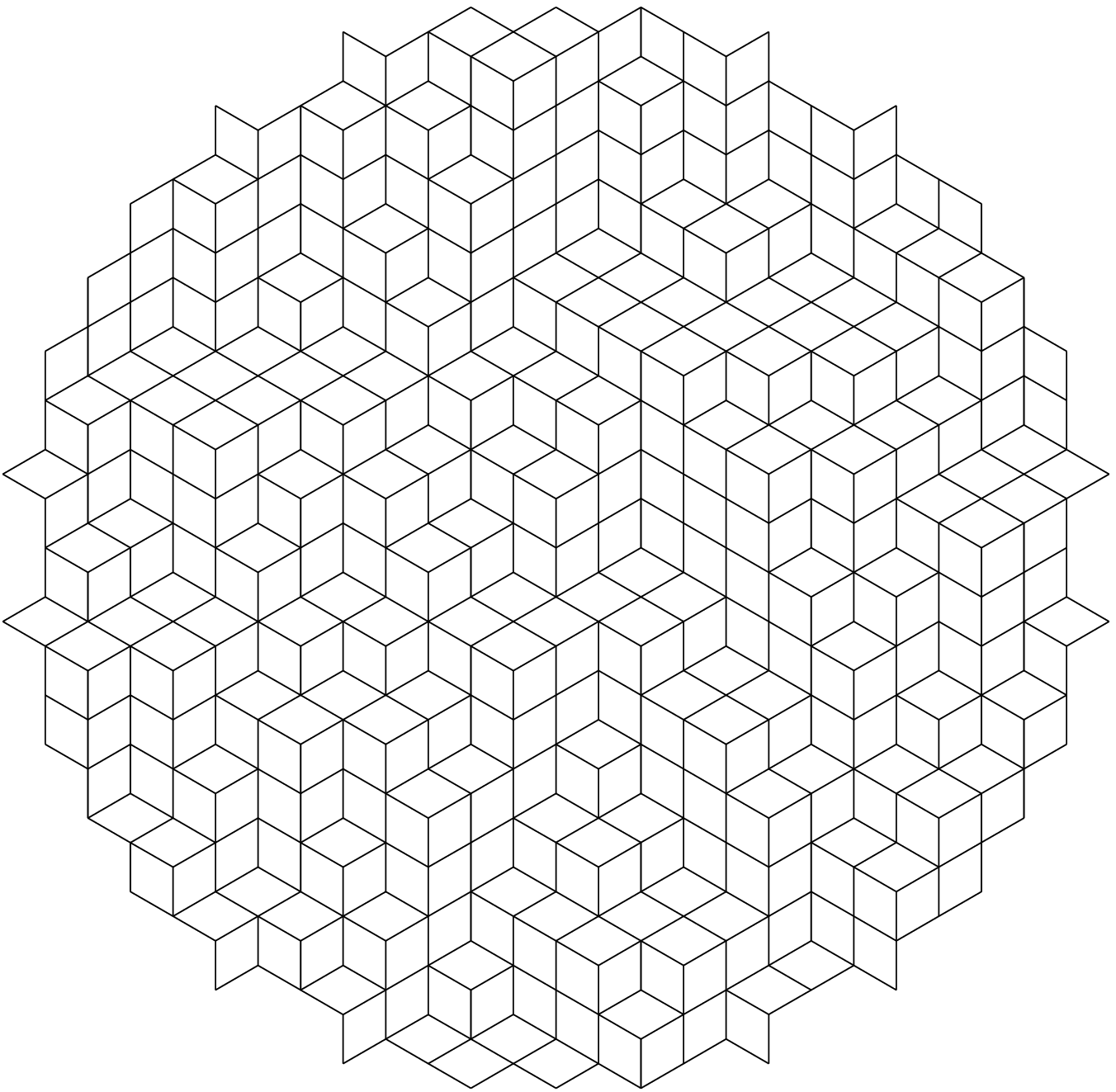,width=3cm}  
\ \psfig{figure=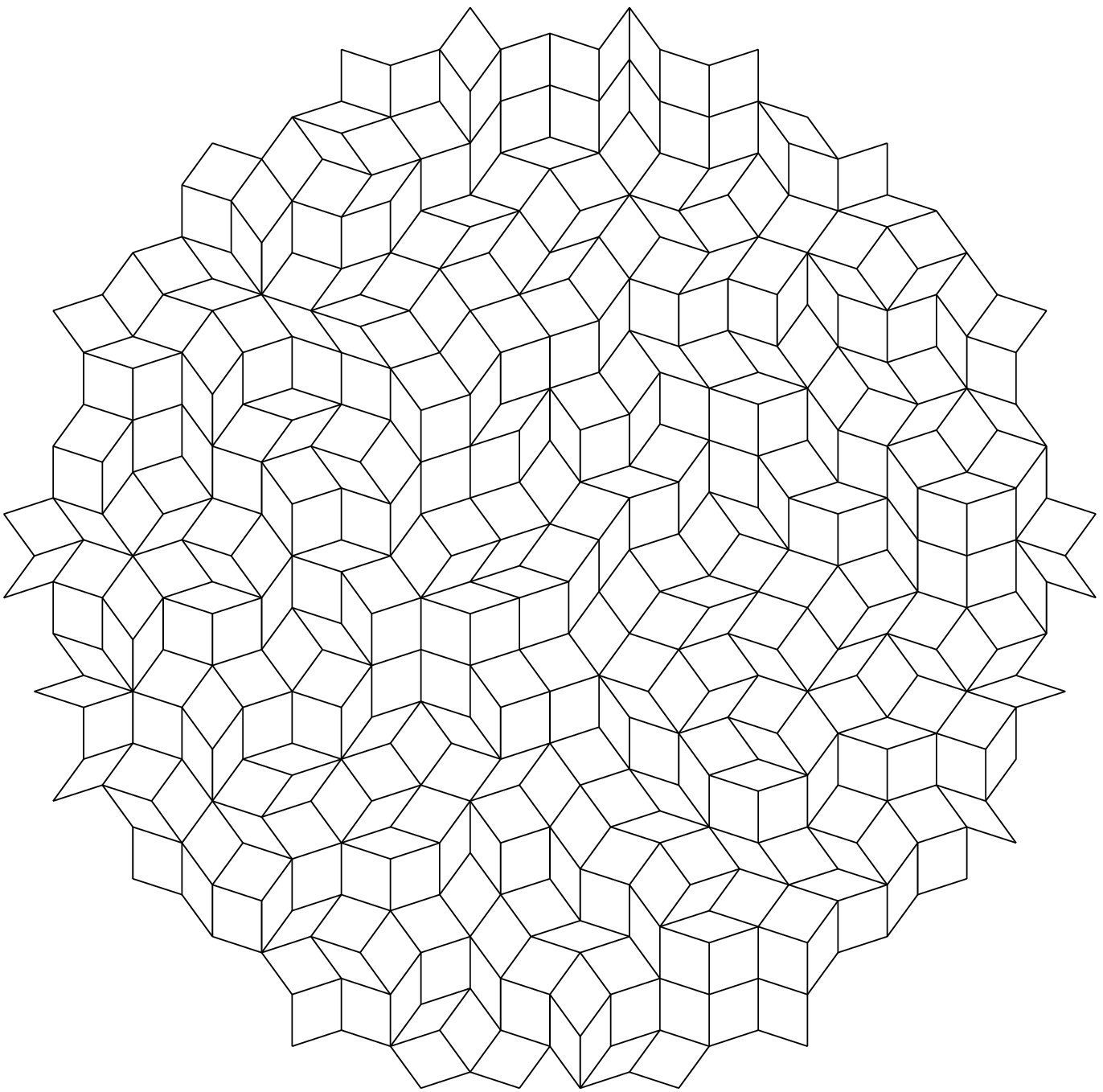,width=3cm} 
\ \psfig{figure=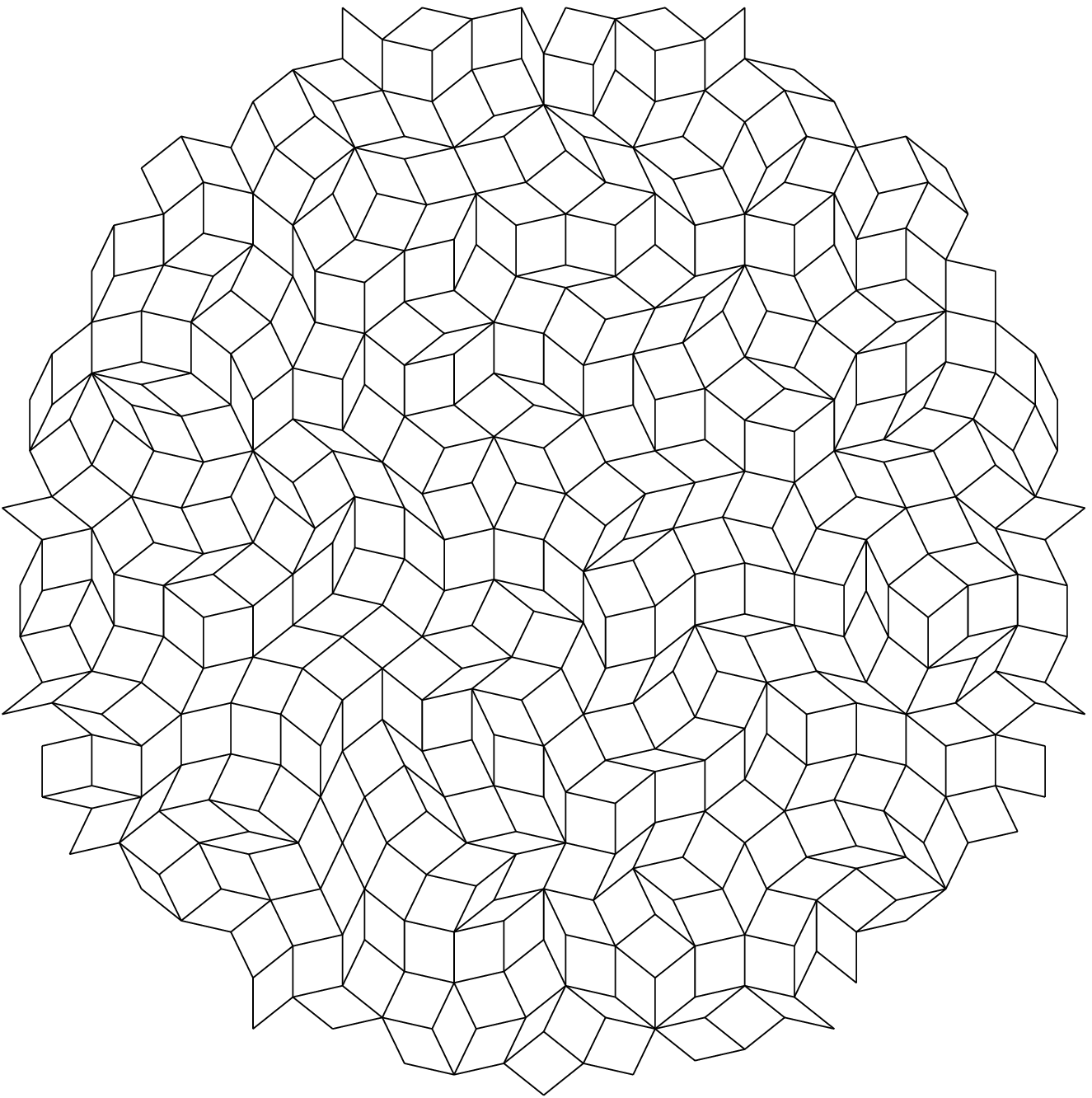,width=3cm} 
\ \psfig{figure=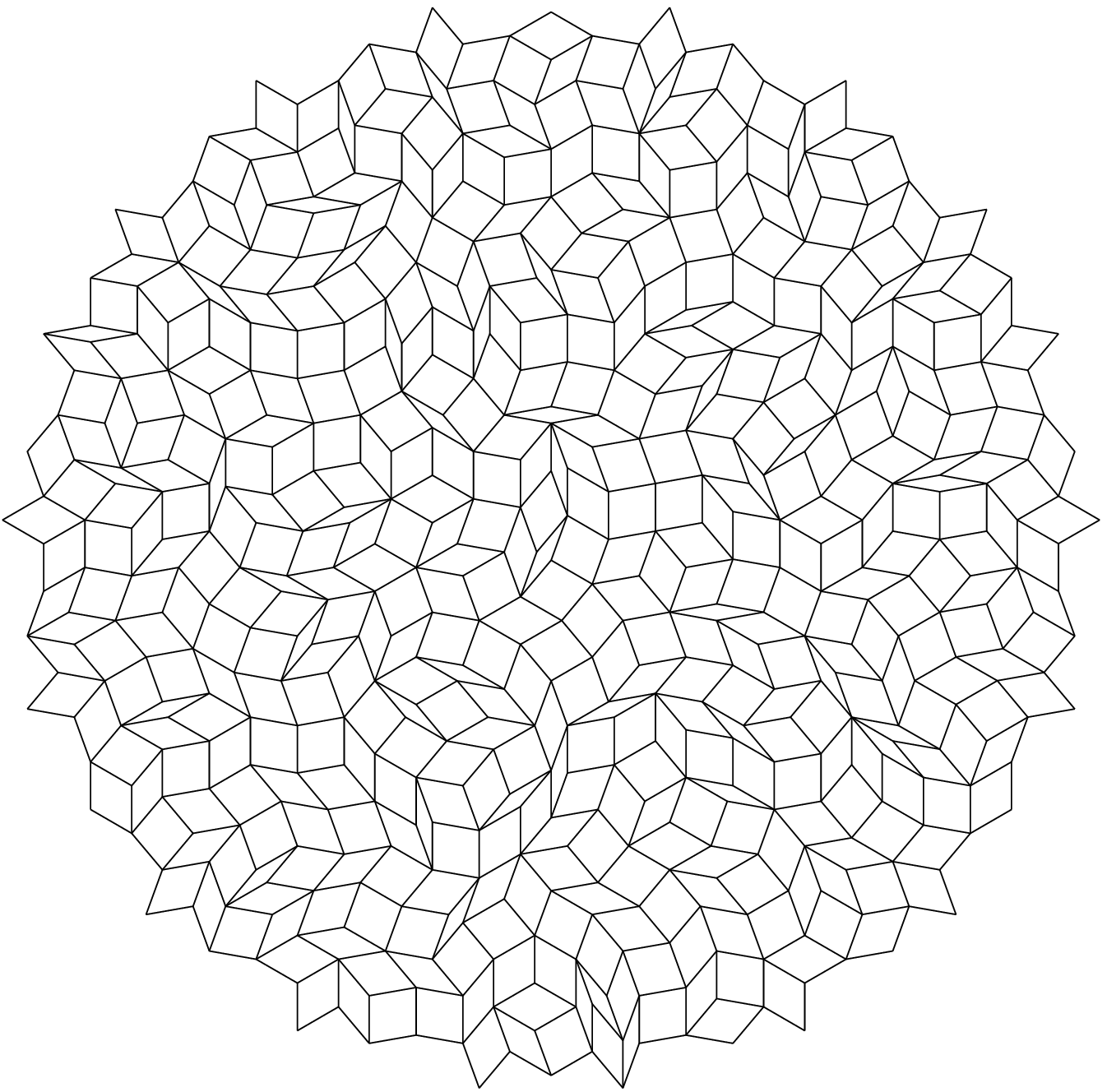,width=3cm} \
\end{center}
\caption{Examples of free boundary $d= 2$ tilings with $D=3,5,7$ and $9$.}
\label{pavagesD2}
\end{figure}

The discovery of quasicrystals by Shechtman, {\em et
al.}~\cite{shechtman}, and the proposal of Levine and
Steinhardt~\cite{Levine} of their possible relationship with Penrose
tilings~\cite{Penrose74}, motivated widespread investigation of
tilings as models for quasicrystal structure. It is presumed that
energetically favored atomic motifs form geometrical tiles, and that
these tiles may be arranged quasiperiodically in space to describe the
quasicrystal structure. We consider tiling models defined as
projections from a higher $D$-dimensional lattice into a lower
$d$-dimensional physical space. For example, 10-fold symmetric tilings
may be projected from $D=5$ into $d=2$, and icosahedral tilings may be
projected from $D=6$ into $d=3$. Figure~\ref{pavagesD2} displays
examples of tilings constructed as projections from $D$-dimensional
simple cubic lattices into $d=2$-dimensional physical space.
The difference between the higher dimension $D$ and the tiling
dimension $d$ is called the {\em codimension}.

An important unsolved problem is understanding the interactions among
tiles that may favor particular tile arrangements over others. The
suggestion by Elser~\cite{elser} that random tilings spontaneously
exhibit quasiperiodicity, created a subfield within the area of tiling
theory~\cite{Henley91}. It was shown~\cite{lj}, for at least one atomistic
model quasicrystal~\cite{lj,lbc}, that quasicrystalline order emerges
with random tiling, rather than Penrose-like, order. The best
description for real quasicrystalline materials remains an open
problem, depending whether one assumes that energetic interactions
favor quasiperiodicity~\cite{energy} or that entropy stabilizes the
quasicrystal structure at high
temperatures~\cite{Henley91,lj,trieste}.

In addition to their role in the theory of quasicrystals, random
tiling models appear in a number of other interesting scientific
contexts. Their combinatorial properties, for example their
relationship to generalized partitions~\cite{Mosseri93B,Bibi97,octo01}
make them interesting models for study within pure
mathematics~\cite{Bjorner93,Kenyon93,Elnitsky97,Bailey97,Cohn98,Cohn01,Cerf01,Latapy02}.
Some random tiling models provide examples of exactly solvable
models~\cite{blote,exact,review} of interest within statistical
mechanics. We note a relationship with algorithms for sorting
lists~\cite{Knuth92,Widom97}. Random tiling models have also been
proposed as models for elastic membranes~\cite{Bibi97,membrane_J}.

\begin{table}[htb]
\caption{$d=2$ strain-free rhombus tiling entropies.}
\begin{tabular}{|c|r|r|r|r|r|}
\hline
D        & 2 & 3     & 4     & 5     & $\infty$\\
\hline
$\sigma$ & 0 & 0.323 & 0.434 & 0.481 & 0.568 \\
\hline
\end{tabular}
\label{tab:d2}
\end{table}

The random tiling theory of quasicrystals focuses on two important
properties: The tiling entropy contributes to the configurational
entropy of the quasicrystal, reducing its free energy and enhancing
thermodynamic stability against other competing phases~\cite{trieste};
Variation of the entropy with average phason strain defines the phason
elastic constants. Table~\ref{tab:d2} summarizes existing data on the
configurational entropies of $d=2$ rhombus tilings, showing a trend of
monotonic increase with $D$. Table~\ref{tab:d3} shows corresponding
data for $d=3$ rhombohedron tilings.

\begin{table}[htb]
\caption{$d=3$ strain-free rhombohedron tiling entropies.}
\begin{tabular}{|c|r|r|r|r|r|}
\hline
D        & 3 & 4     & 5      & 6    & $\infty$ \\
\hline
$\sigma$ & 0 & 0.214  & NA     & 0.24 & $<0.323$ \\
\hline
\end{tabular}
\label{tab:d3}
\end{table}

Techniques employed in theoretical studies include analytic exact
solution~\cite{exact,review}, Monte Carlo computer simulation~\cite{mc} and
numerical transfer matrix methods~\cite{tm}. While these methods prove
successful at calculating quantitative values for entropy and elastic
constants, they are not always easy to implement, and they often do
not enhance our intuitive and qualitative understanding of the
problem. The technique of mean-field theory~\cite{MFTref} in
statistical mechanics provides an approximation that can be analytically
simple to implement and can provide a direct link between quantitative
calculation and our qualitative picture of the physics of the
problem. Mean-field theory often becomes exact in certain limits, such as
the limit of infinite dimension. A mean-field theory for random
tilings has never been precisely defined.

The search for an analogue of mean-field theory applicable to random
tilings motivates our study of random tiling models with high
rotational symmetry.  We anticipate~\cite{Henley91} that the infinite
$D$ limit may prove easier to analyze than specific finite values of
$D$.  Experience gained in the study of this limit may later be
transfered back to the finite $D$ values of greatest physical
interest.  Furthermore, a study of $D$ dependence of entropy or
elastic constants might in itself yield qualitative understanding of
the values of entropy and elastic constants for small $D$. Other
researchers~\cite{Korepin,Socolar,Mermin} have studied deterministic
quasiperiodic tilings of arbitrary rotational symmetry, but have
considered neither the high symmetry limit nor the case of random
tilings.

We mention, in addition, that high symmetry tilings are good
candidates of models for topologically disordered graphs, different
from the usual random graphs~\cite{Bollobas} because the coordination
number of a vertex is a random variable, with algebraically decreasing
long-range correlations.  Quantum diffusion on such
tilings has recently been studied by Vidal {\em et al.} and was
found to exhibit an interesting diffusive behavior~\cite{Vidal03}.
These tilings could even prove interesting as models for structural glasses.

A word of caution should be given on the title referring to ``high
symmetry'' tilings. In two dimensions, this statement is unambiguous
owing to the possibility of generating tilings whose Fourier spectrum
will display point symmetries of arbitrary order. In three dimensions
however, the highest point group symmetry, the icosahedral one of
order 120, is already reached with codimension 3 tilings. ``High
symmetry'', in three and higher dimensions, has therefore a less
formal meaning in our description. For a given tiling in dimension
$d$, we can map the whole set of edge orientations onto points on a
$(d-1)$-dimensional hypersphere. As $D$ increases, these
representative points cover more and more densely the hypersphere,
approaching, but never meeting, the highest continuous point symmetry
of the hypersphere itself. This is what is meant by ``high symmetry'' in
the title.

Two previous papers~\cite{Widom97,Bibi00} presented initial studies of
this problem. The first paper proposed an upper bound on the entropy
in the limit of large $D$ and discussed problems associated with the
thermodynamic limit of tiling models. The second one~\cite{Bibi00}
presented a preliminary mean-field approach of the entropy
calculation. In the present paper we explain the previous work in
greater detail. In addition, we inspect spatial correlations and
phason elasticity. All tilings considered in this paper are free
boundary tilings (see section~\ref{defs}). Fixed boundary tilings are
analyzed by computer simulation in the following paper~\cite{paperII}.
Table~\ref{estimate} summarizes the available estimates of the
large $D$ entropy.

\begin{table}[htb]
\caption{Entropy estimates in the large $D$ limit.}
\begin{center}
\begin{tabular}{|l|l|l|}
\hline
Value  & Method            & Reference \\
\hline
0.998  & Rigorous upper bound & [\cite{Knuth92}] \\
0.693  & Conjectured upper bound upper bound       & eq.~(\ref{eq:our_bound}) \\
0.6    & Extrapolated exact values & Ref. [\cite{Widom97}] \\
0.598  & Mean-field theory & eq.~(\ref{ent_mft}) \\
0.568  & Monte Carlo simulation        & Ref. [\cite{paperII}] \\
0.231  & Rigorous lower bound & [\cite{Knuth92}] \\
\hline
\end{tabular}
\end{center}
\label{estimate}
\end{table}

The organization of this paper is as follows: We begin in
section~\ref{defs} with definitions and concepts that apply generally
to random tiling models.  Then, in section~\ref{MFT} we describe our
mean-field theory of two-dimensional tilings of high rotational
symmetry. We start with an upper bound (section~\ref{order0}) then a
more accurate estimate (section~\ref{MFcalc}) based on path counting
arguments. We investigate the nature of spatial correlations in
section~\ref{correl} followed by consideration on finite $D$
corrections. Section~\ref{interactions} tackles de Bruijn line
interactions and phason elasticity. Tilings of dimension $d\ne 2$ are
examined in section~\ref{dnot2}.

\section{Characterization of tilings}
\label{defs}
\setcounter{equation}{0}

In this section, we introduce the basic material used throughout the
paper to characterize and manipulate two-dimensional random tilings.

A tiling is a covering, without gaps or overlaps, of a given region of
a $d$-dimensional Euclidean space.  In the present paper, the tiles
are $d$-dimensional rhombohedra, which we will generically call
``rhombi'' in the following.

Tiling systems can have different boundary conditions, such as free,
fixed or periodic ones. For example, the tilings in
figure~\ref{pavagesD2} illustrate free boundary tilings of various
rotational symmetries. The free boundary thermodynamic limit is taken as
follows: Consider all tilings that cover a circular region of space
${\cal R}$ (whose area will tend towards infinity). For a given ${\cal
R}$, tilings contain fixed numbers of tiles of each type, with
relative frequencies governed by the average composition of the tiling
we wish to study. The total number of tiles is chosen so they
encompass a total area slightly larger than area of the region ${\cal
R}$ that must be covered. Every tile must intersect ${\cal R}$ or
share an edge with a tile that does so. This ensures a reasonably
compact tiling and gives us needed control over the composition.

It is generally supposed that free and periodic boundary tilings have
the same entropies (provided their compositions match) at the
thermodynamic limit. But fixed boundary tilings display a less usual
behavior: such boundaries have such a macroscopic effect on tilings
that their entropy per tile at the infinite size limit can be strictly
smaller than the free boundary one. For example, in the case of
``hexagonal'' tilings of 60$^{\circ}$ rhombi ($D=3$ and $d=2$), when
all orientations of rhombi occur with equal probability, the fixed
boundary entropy per tile $\sigma_{fixed}=0.261$~(\cite{elsershape})
while the free boundary entropy per tile
$\sigma_{free}=0.323$~(\cite{Wannier}). In addition to a qualitative
argument by Elser~\cite{elsershape}, this phenomenon has more recently been
rigorously explained and described~\cite{Bibi97B,Cohn98,Cohn01}: the
local entropy density can be calculated in any point of the tiling; it
displays a gradient between the regions near the boundary, where the
entropy is vanishing, and the central region, where the entropy is a
free boundary one. In other words, it is only at the very center of
the tiling that the tiling loses the influence of the boundary.

\subsection{Membrane representation of tilings}

Here, we briefly recall how tilings are conveniently coded by
(mono-valued) functions from the ``physical real space'' to the
``perpendicular'' space.

Rhombic tiles can be considered as the projections onto a
$d$-dimensional space of the $d$-dimensional faces of a hypercube in a
space of higher dimension $D$. Conversely, any rhombic tiling in the
$d$-dimensional space can be ``lifted'' to a $d$-dimensional
continuous membrane embedded in the $\Z^D$ hypercubic lattice. When
this membrane is projected back into $d$-dimensional space, the facets
of the $D$-dimensional lattice project precisely onto the
$d$-dimensional rhombic tiles.  Such a membrane is said to be {\em
continuous} and {\em directed} to emphasize the fact that, when
projected, is does not exhibit any gaps or overlaps. We say that we
are dealing with $D \ra d$ membranes or tilings. This {\em membrane}
representation of rhombus
tilings~\cite{Henley91,Bibi97,membrane_J,Bibi97B} is a useful tool in
quasicrystal science.

Notation related to the $D$-dimensional geometry and projection is
described in appendix~\ref{hyper}. Let $(\eh_i)_{i=1,\ldots,D}$ denote
the basis of the $\Z^D$ hypercubic lattice and denote by
$\eh^\parallel_i$ the normalized projection of $\eh_i$ on the
$d$-dimensional space where tilings lie. The projection direction is
(arbitrarily) chosen so that the basis vectors $\eh^\parallel_i$ all
have equal length and angles that are multiples of $\pi / D$ (see
appendix~\ref{hyper}). We call this collection of vectors a {\em
regular fan}.

A rhombic tile has $d$ edge orientations
$\{\eh^\parallel_i,\eh^\parallel_j, \cdots, \eh^\parallel_k\}$ with
$(1 \le i<j< \cdots < k \le D)$. We denote this tile species as $T_{ij
\cdots k}$.  Note that there are $\simp Dd$ different species of
tiles. Examples of $D \ra 2$ tilings are displayed in
figure~\ref{pavagesD2}. The first one can be seen as a membrane
embedded in a 3-dimensional cubic lattice, as may be more clear in the
following figure~\ref{terraces}.

\begin{figure}[ht]
\begin{center}
\ \psfig{figure=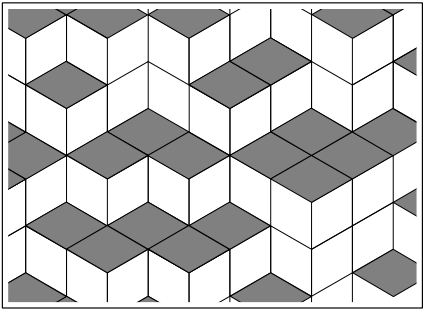,height=5.5cm} \
\end{center}
\vspace{-5mm}
\caption{Three-dimensional image effect. Gray rhombi can be
seen as horizontal terraces.}
\label{terraces}
\end{figure}

\begin{figure}[ht]
\begin{center}
\begin{tabular}{ccc}
\parbox{3in}{\vfill \psfig{figure=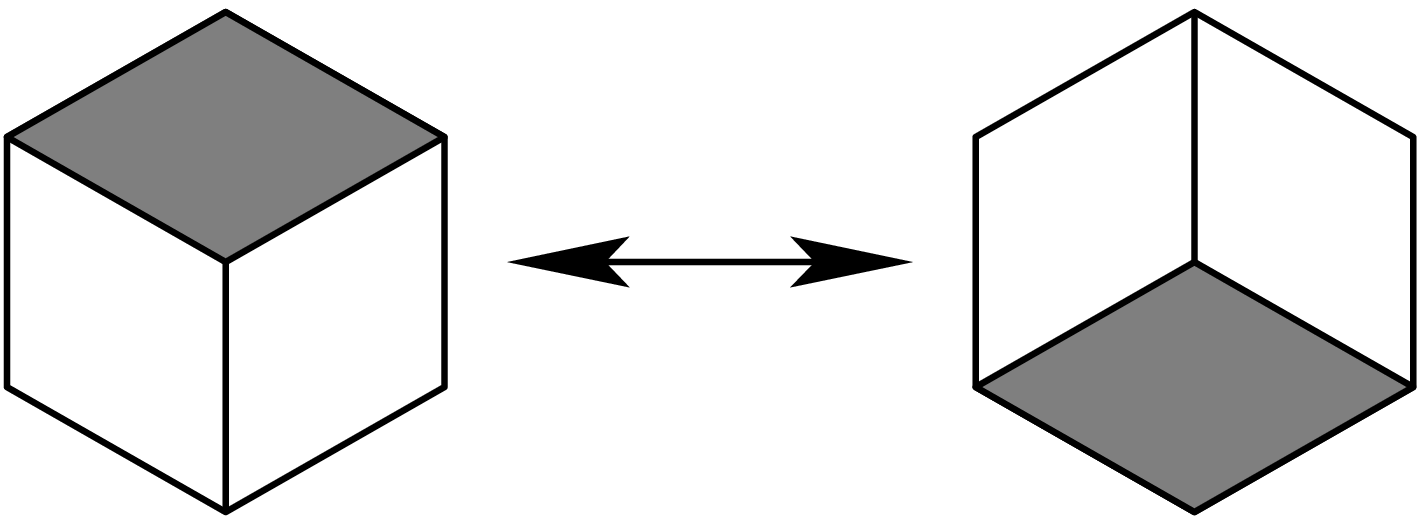,height=1.6cm} \vfill} & \phantom{AAAA}
\parbox{3in}{\vfill \psfig{figure=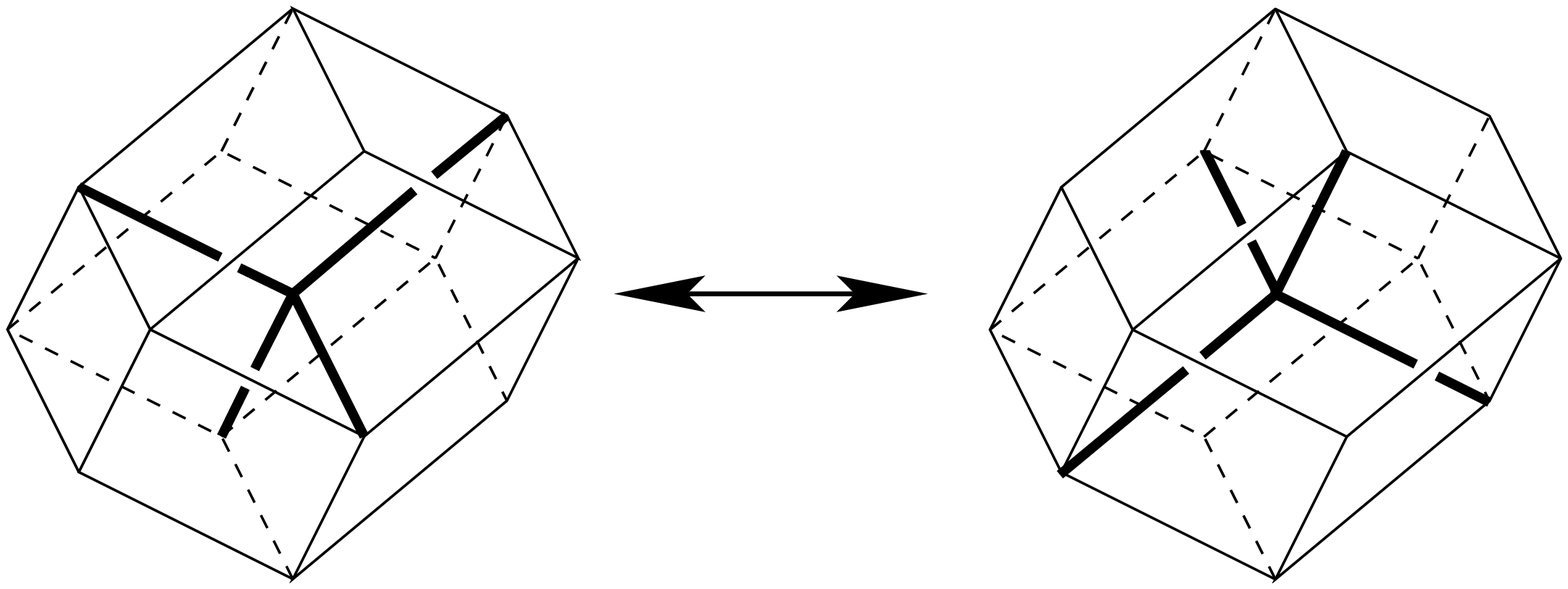,height=28mm} \vfill}    \\
  \end{tabular}
\end{center}

\caption{\label{flips} Two examples of elementary flips in
  2 and 3 dimensions, involving respectively 3 and 4 rhombic tiles.}
\end{figure}

Such random tilings have internal degrees of freedom, the so-called
{\em localized phason flips}, which consist of local rearrangements of
tiles. Such local flips are displayed in figure~\ref{flips}. In $d$
dimensions, a flip involves $d+1$ tiles that fill a small
zonotope~\cite{Coxeter}.  Note that in 2 dimensions, the ergodicity of
tiling ensembles {\em via} elementary flips is
proven~\cite{octo01,Kenyon93}, whereas it is an open question in 3 and
more dimensions, even if some significant progress has been achieved
recently~\cite{Desoutter04}. This local degree of freedom is the
source of configurational entropy in random tilings.

A functional representation of these directed membranes will prove
useful. Denote the $d$-dimensional ``physical'' or ``real'' space
containing the actual tiles by $\EC^\parallel=\R^d$, and the
perpendicular $(D-d)$-dimensional space by $\EC^\perp=\R^{D-d}$. Since
membranes representing tilings are directed, they are the graphs of
single-valued continuous functions $h : \EC^\parallel \ra
\EC^\perp$. Such functions $h$ are called the {\em height functions}
of the corresponding tilings. The detailed construction of the spaces
$\EC^\parallel$ and $\EC^\perp$ is described in appendix~\ref{hyper}
when $d=2$. Executing an elementary flip of vertex $v$ changes the
value of the height function locally at vertex $v$.

Since the membranes are embedded in a hypercubic lattice, they are
irregular and faceted at on the length scale of individual tiles. The
local height function $h$ is usually coarse-grained to get a smoother
function $\bar{h}$ which closely tracks the local function $h$ over
long distances but which is nearly insensitive to individual single
vertex flips~\cite{Henley91,Bibi97,lpw}. The function $\bar{h}$
describes a smoothed membrane. Then one defines the {\em phason
gradient} $\vect{E}=\nabla \bar{h}$.  This $(D-d)\times d$ dimensional
tensor controls the fractions of the different species of tiles.

The entropy can be written as a functional of the phason
gradient~\cite{Henley91}.  Moreover, the random tiling model
hypothesis states that this latter entropy has a unique maximum,
corresponding to tile fractions maximizing the symmetry. The
orientation of the real space $\EC$ is chosen so that the gradient is
zero at this maximum, and the model states that the entropy density has a
quadratic behavior near this maximum:
\begin{equation}
\sigma = \sigma_{\mbox{\scriptsize max}} - {1 \over 2} \; \nabla
\bar{h} \cdot {\bf K} \cdot \nabla \bar{h} + o(|\nabla \bar{h}|^2).
\label{hypothesis}
\end{equation}
By analogy with an usual elastic theory, the tensor {\bf K} is called the
tensor of {\em phason elastic constants}.

The best known rhombus tilings of the plane is the
Penrose~\cite{Penrose74} tiling, which displays an exact
quasiperiodicity. The membrane representation of a Penrose tiling is a
flat, horizontal, sheet. The phason strain of this tiling vanishes. In
addition, the Penrose tiling is virtually unique (up to translations)
because it forbids localized phason flips. The random tiling
hypothesis implies that the most typical tilings in a random tiling
ensemble mimic the long-range order of the Penrose tiling, while
enjoying the entropy of short-range fluctuations.

Random and perfect tilings with vanishing phason strain exhibit the
highest symmetry consistent with the projection from $D \ra d$. In two
dimensions, this symmetry includes~\cite{Mermin} $2D$-fold rotational
symmetry. A given tiling itself generally has no point about which it
is invariant under rotations, but two-point and higher correlation
functions do exhibit the symmetry. In addition, there are two families
of mirror symmetries, denoted ``along'' and ``between'' according to
the relation between the mirror plane and the basis vectors
$\eh^\parallel_i$.

\subsection{De Bruijn dualization and iterative construction of tilings}
\label{itere}

The de Bruijn dualization~\cite{debruijn} provides another
representation of random tilings. We present this technique in two
dimensions and generalize it to higher dimensions when needed.

\begin{figure}[ht]
\begin{center}
\ \psfig{figure=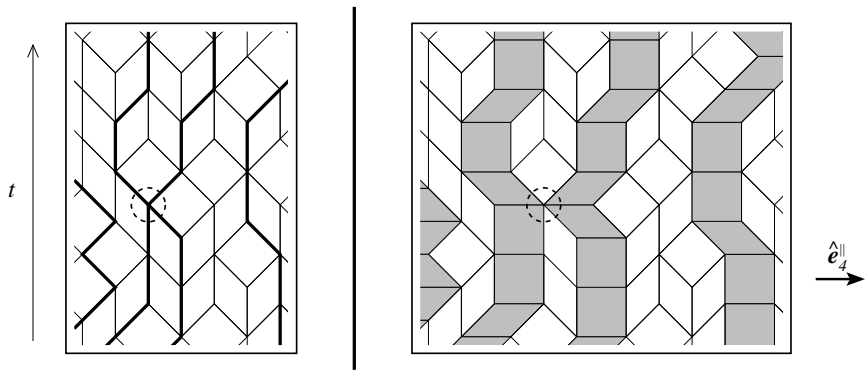,height=5.5cm} \
\end{center}
\caption{Iterative process for the construction of $D \ra 2$ tilings. Left
shows paths on a re-oriented $3 \ra 2$ tiling. Right shows the
resulting $4 \ra 2$ tiling.  A non-crossing contact between two de
Bruijn lines of the same family is emphasized.}
\label{contact}
\end{figure}

De Bruijn grids~\cite{debruijn,Socolar85,Gahler86} are made up of
lines, the de Bruijn lines, that are sometimes called {\em
worms}. These lines are made of adjacent tiles sharing an edge of
given orientation. It is always possible to extend these lines through
the whole tiling up to a boundary tile. An example is presented in
figure~\ref{contact}. Since there are $D$ different edge orientations,
there are $D$ different de Bruijn families.  A rhombic tile
corresponds to the intersection of two lines of different
families. Hence there are $\simp D2 = D(D-1)/2$ tile species.  There
are no triple intersections, and lines of a same family never
intersect, even in an infinite tiling, because no rhombus has four
parallel edges.

This de Bruijn representation of tilings suggests an iterative
construction of $D \ra 2$ tilings in terms of directed random walkers
on $D-1 \ra 2$ tilings.  The basic idea is illustrated in
figure~\ref{contact}. A $D-1 \ra 2$ tiling is sheared to make room for
a new de Bruijn line family by taking the vectors $\eh^\parallel_i$
that initially lie at angles ${\pi i} / (D-1)$ and replacing them
with angles ${\pi i} / D$ (see Fig.~\ref{contact}, left).  Next,
paths are chosen on the $D-1 \ra 2$ tiling (dark lines in
Fig.~\ref{contact}, left).  They go from bottom to top, as symbolized
by a ``time arrow'' $t$. These paths must be ``opened'' in the
direction $\eh_D^\parallel$ in order to form de Bruijn lines (shaded
in figure \ref{contact} right) of the $D$-th family and therefore a $D
\ra 2$ tiling. These paths are directed in the positive time direction
and they do not intersect within the same family. The figure shows a
contact among paths that is not an intersection.  Conversely, de
Bruijn lines of family $D$ in a $D \ra 2$ tiling may be collapsed to
directed walks in a $D-1 \ra 2$ tiling. There is thus a one-to-one
correspondence between $D \ra 2$ tilings with $p$ de Bruijn lines of
family $D$ and collections of $p$ non-intersecting walks on $D-1 \ra
2$ tilings.

\subsection{Grid parameters, phason strain and tile fractions}
\label{tile_fractions}

In this subsection, we introduce several macroscopic quantities that
characterize a two-dimensional random tiling. Apart from the phason
gradient $\vect{E}$ previously defined, a tiling can also be
characterized at large scale either by the tile fractions or by
geometric parameters. This subsection also provides the relationships
between these quantities (\ref{relations}) as well as useful relations
that constrain them (\ref{misc}).

\subsubsection{Definitions}

\begin{figure}[ht]
\begin{center}
\ \psfig{figure=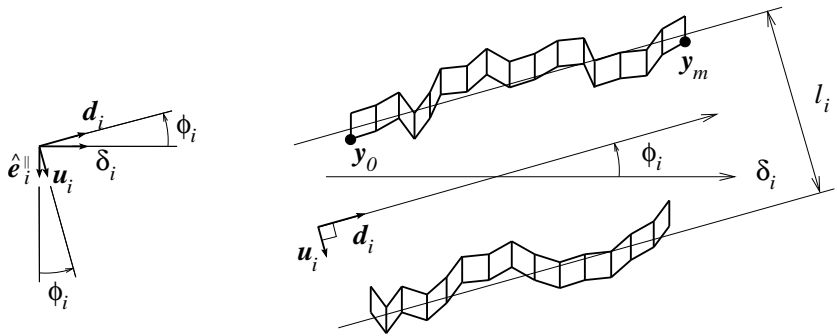,width=11cm} \
\end{center}
\caption{Definition of quantities $l_i$~-- the mean distance between
two neighbor lines of the $i$-th de Bruijn family~-- and $\phi_i$~--
the mean angle that the lines of this family make with \mbox{\boldmath
$\delta$}$_i$.  The vector $\vect{u}_i$ is given by the vector
$\eh_i^\parallel$ rotated by the angle $\phi_i$. The orientation
$\vect{d}_i$ of a line of the family $i$ is given by the average of the
vectors $\fg_j$ which define its tiles.}
\label{2worms}
\end{figure}

We now address two macroscopic characterizations of de Bruijn lines,
the mean line spacings and orientations, and link these to the phason
strain ${\bf E}$ and the tile fractions $n_{ij}$ (the fraction of
tiles that are of type $T_{ij}$). De Bruijn lines of family 
$i$ are characterized by the mean spacing $l_i$ between two lines of
the family $i$ and the mean angle $\phi_i$ that lines of this family
make with their nominal direction . The nominal direction (denoted by
$\mbox{\boldmath $\delta$}_i$) is normal to $\eh_i^\parallel$ (see
figure~\ref{2worms}). For a maximally symmetric tiling with vanishing
phason strain ${\bf E}=0$, rotational symmetry guarantees that the
mean line spacing $l_i$ is independent of the line family $i$. We
shall calculate its common value shortly. The strain-free tiling's
mirror symmetries guarantee that each angle $\phi_i=0$. Deviations of
$l_i$ and $\phi_i$ from their strain-free values control nonzero
values of ${\bf E}$ and determine the tile fractions $n_{ij}$.
By definition, $l_i > 0$ and $- \pi/2 < \phi_i < \pi/2$.

Figure~\ref{2worms} represents lines of the $i$-th family as sequences
of tiles $T_{ij}$, each tile defined by the vector $\eh_i^\parallel$ and
some other vector $\eh_j^\parallel$. Let $\vect{y}_0$ be the relative position
of a given tile in a de Bruijn line, and follow this line in the
plane. The position $\vect{y}_m-\vect{y}_0$ of the $m$-th tile is the
sum of the vectors defined by the successive edges of the tiles $T_{ij}$ of the
line between tile 0 and tile $m$.  These vectors are {\em a
priori} equal to $\pm \eh_j^\parallel$. Therefore we define $\fg_j=\pm
\eh_j^\parallel$ where the sign is chosen so that $\fg_j \cdot
\mbox{\boldmath $\delta$}_i > 0$. For large $m$, the number of
tiles $T_{ij}$ on any typical line is proportional to the tile
fraction $n_{ij}$. The mean direction of this line is therefore
\begin{equation}
\vect{d}_i=\sum_{j=1 \atop j \neq i}^D n_{ij} \fg_j.
\label{d_alpha}
\label{direction}
\end{equation}

In general, a random tiling ensemble is completely characterized by
$\{l_i\}$ and $\{\phi_i\}$. Since $D-2$ families of lines must be
added to an original square lattice ($D=2$) to define a $D \ra 2$
tiling we need $2(D-2)$ independent macroscopic parameters to
characterize a tiling: $l_i$ and $\phi_i$ for $i=3, 4, \dots, D$.
Note that $2(D-2)$ is precisely the number of coefficients of the
global phason gradient $\vect{E}=\nabla \bar{h}$. By comparison, in
reference [\cite{lpw}], $4 \ra 2$ tilings are characterized by 6 tile
concentrations $d_{ij}$, and 2 exact relations constrain
them. Therefore these tilings are characterized by 4 independent
parameters, as predicted above.

\subsubsection{Relations between macroscopic parameters}
\label{relations}

Consider now the tile fractions $n_{ij}$ and their relation to $\{l_i\}$
and $\{\phi_i\}$.  The tile number fractions $n_{ij}$ are proportional
to their density $d_{ij}$ (number per unit area) {\em via}
\begin{equation}
\label{dijnij}
n_{ij}={{d_{ij}}\over{\sum_{k<l} d_{kl}}}.
\end{equation}
Conversely, 
\begin{equation}
\label{nijdij}
d_{ij}={{n_{ij}}\over{\sum_{k<l} n_{kl} \sin |\pi(l-k)/D|}}
\end{equation}
because the area of a tile $T_{ij}$ equals $\sin |\theta^*_{ij}|$
where $\theta^*_{ij} = \pi(j-i)/D$ is the oriented angle between
$\eh^\parallel_i$ and $\eh^\parallel_j$. 

Note that $\theta^*_{ij}$ is also the angle between de Bruijn line
directions $i$ and $j$ in a strain-free tiling. Generally speaking,
throughout the text, starred quantities will refer to unstrained
tilings with $\vect{E}=0$.

Relation~(\ref{dijnij}) is useful because the tile densities are given in a
simple fashion by~\cite{lpw,Henley88}
\begin{equation}
d_{ij}=| \vect{m}_i \times \vect{m}_j |,
\label{area_frac}
\end{equation}
where $\vect{m}_i$ (see appendix~\ref{hyper}) is the parallel space
gradient of the $i^{th}$ component of the smoothed membrane hyperspace
coordinates.  We also define a vector $\vect{u}_i$ for each line family:
it is the unit vector along the shortest direction from one
line of the family to the next; it is perpendicular to $\vect{d}_i$.
Examining figure~\ref{2worms}, if one travels a distance
$l_i$ in the direction $\vect{u}_i$, then one crosses one de Bruijn
line of the family $i$ on average and the $i^{th}$ component of the
hyperspace coordinates increases by one on average. In contrast, if
one travels any distance in the orthogonal direction ${\bf d}_i$, then
the $i^{th}$ component of the hyperspace coordinates remains constant.
Thus
\begin{equation}
\vect{m}_i = {1 \over l_i} \; \vect{u}_i
\end{equation}
and equation~(\ref{area_frac}) becomes
\begin{equation}
\label{dij}
d_{ij} = {1 \over l_i l_j} | \vect{u}_i \times \vect{u}_j |
= {1 \over l_i l_j} | \sin \theta_{ij} |,
\end{equation}
where the angle between de Bruijn line directions
\begin{equation}
\label{thetaij}
\theta_{ij}=\theta^*_{ij}+\phi_j-\phi_i.
\end{equation}

Next we relate the distances $\{l_i\}$ and angles $\{\phi_i\}$ to the
phason strain $\vect{E}$. Owing to relation~(\ref{E_and_mi}) of
appendix~\ref{hyper}, we have
\begin{equation}
\vect{E} = \sum_{i=1}^D {1 \over l_i} \; \eh^\perp_i \otimes
\vect{u}_i.
\label{E_li_phii}
\end{equation}
Defining $\vect{E}^2$ as
the sum of the squares of the components of $\vect{E}$, then
\begin{equation}
\vect{E}^2 = \sum_{i=1}^D \sum_{j=1}^D {1 \over l_i l_j} 
(\vect{u}_i \cdot \vect{u}_j) (\eh_i^\perp \cdot \eh_j^\perp).  
\end{equation}
We prove in appendix~\ref{hyper} that if $i \neq j$, then
$\eh_i^\perp \cdot \eh_j^\perp = -2 \cos \theta^*_{ij}/(D-2)$.
Hence
\begin{equation}
\vect{E}^2 = - {2 \over D-2} \sum_{i \neq j} {1 \over l_i l_j} 
\cos \theta_{ij} \cos \theta^*_{ij} + \sum_{i} {1 \over l_i^2}.
\label{E_li_phii2}
\end{equation}

To determine the tile fractions $n^*_{ij}$ and the mean line spacing
$l^*_i$ for strain-free tilings (we recall that starred quantities
refer to unstrained tilings), remind that the tiling symmetry
guarantees $l_i$ is independent of $i$, and $\phi_i=0$. Inspecting
eqs.~(\ref{dij}) and~(\ref{dijnij}), we may write
$n_{ij}=C|\sin\theta^*_{ij}|$. To determine the normalization constant
$C$, note that
\begin{equation}
1 \equiv \sum_{i<j} n_{ij}^* = C  \sum_{0\leq i<j <D} \sin \pi {j-i \over D}.
\end{equation}
In the limit of large $D$ one then finds $C \simeq \pi/D^2$, so that
\begin{equation}
\label{nijstar}
n_{ij}^* = {\pi \over D^2} |\sin \theta_{ij}^*|.
\end{equation}
The fraction $n_i$ of tiles belonging to a given family $i$ of lines,
in a maximally symmetric tiling at large $D$,
\begin{equation}
\label{meanni}
n_i^* = \sum_{j \neq i} n_{ij}^* \simeq {\pi \over D^2} {2D \over \pi}
= {2 \over D}
\end{equation}
allows us to compute the mean distance between de Bruijn lines of a
given family. Since tiles of family $i$ have, on average, the same
area as tiles of the whole tiling, the above {\em numerical} tile
fraction is also an {\em area} tile fraction. Since a fraction $2/D$
of tiles belong to lines of a family and the width of such a line is
1, the mean distance between lines is
\begin{equation}
\label{li}
l_i^* = D/2.
\end{equation}
Setting $l_i = l_i^*$ and $\theta_{ij}=\theta_{ij}^*$ for all $i$ and
$j$ in (\ref{E_li_phii2}), we find again in a consistent way
$\vect{E}^2=0$. Moreover, eq.~(\ref{dij}) implies 
\begin{equation}
d_{ij}^* = \frac{4}{D^2} |\sin \theta_{ij}^*| = \frac{4}{\pi} \; n_{ij}^*
\label{dijstar}
\end{equation}
in the large $D$ limit.

\subsubsection{Miscellaneous constraints}
\label{misc}

\begin{figure}[ht]
\begin{center}
\ \psfig{figure=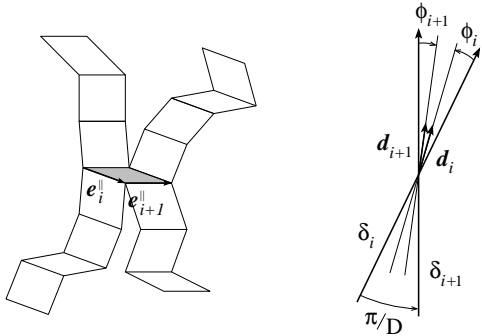,height=4.5cm} \
\end{center}
\caption{Left: Intersection of two lines of two consecutive families
$i$ and $i+1$. Right: The corresponding directions
$\vect{\delta}_i$, $\vect{\delta}_{i+1}$ (outer thick lines) and
$\vect{d}_i$, $\vect{d}_{i+1}$ (inner thin lines).
The vectors $\vect{d}_i$ and $\vect{d}_{i+1}$ are constrained by the
condition $(\vect{d}_i \vect{d}_{i+1})>0$, which constrains in turn
the angles $\phi_i$ and $\phi_{i+1}$.}
\label{intersection}
\end{figure}

Geometrical considerations constrain the parameters $\{l_i\}$ and
$\{\phi_i\}$. We describe a few such constraints here. The condition
\begin{equation}
\label{condition:phi:i}
\phi_{i-1} - {\pi \over D} \leq \phi_i \leq \phi_{i+1} + {\pi \over D}
\end{equation}
imposes a certain regularity to the angles $\phi_i$ as a function of
$i$. This constraint holds because lines of two consecutive families $i$ and
$i+1$ intersect only at a tile $T_{i,i+1}$, with edges $\eh_i^{\parallel}$ and
$\eh_{i+1}^{\parallel}$ as displayed in figure~\ref{intersection}.
The oriented angles $(\vect{d}_i \vect{d}_{i+1})$ and
$(\vect{\delta}_i \vect{\delta}_{i+1})$ have the same sign. Indeed,
were these angles of opposite sign, no tile type
could match at their intersection. As a consequence $(\vect{d}_i 
\vect{d}_{i+1}) = (\vect{\delta}_i \vect{\delta}_{i+1}) - \phi_i +
\phi_{i+1} \geq 0 $. Owing to $(\vect{\delta}_i \vect{\delta}_{i+1}) =
\pi/D$, one gets $ \phi_{i+1} - \phi_i \geq -\pi/D$ and condition
(\ref{condition:phi:i}) (see figure~\ref{intersection}). Such
conditions mean that at large $D$, when one adds a new family of lines
in the iterative process, the angle $\phi_D$ is highly constrained by
the pre-existing underlying tiling. For example, if a family $i=D$ is
added in a strain-free tiling where $\phi_1=\phi_{D-1}=0$, the
condition~(\ref{condition:phi:i}) reads $-\pi/D \leq \phi_D \leq \pi/D$.

Since amongst the $2D$ parameters $\{l_i\}$ and $\{\phi_i\}$, only
$2(D-2)$ are independent, there exist 4 relations between
them. Of these four we will need this identity
\begin{equation}
\label{normalize}
\sum_{i<j} {1 \over l_i l_j} |\sin \theta_{ij}| |\sin \theta_{ij}^*| = 1
\end{equation}
which reflects the fact that the total contribution of tile areas per
unit area is by definition equal to 1. In other words, $\sum_{i<j}
d_{ij} |\sin\theta_{ij}^*| =1$ where $d_{ij}$ is given by
eq.~(\ref{dij}). This relation can also be derived from
eq.~(\ref{nijdij}).

In the large $D$ limit we have infinitely many tile species, so we
expect that all $n_{ij} \ra 0$.  When taking the large $D$ limit we
wish to ensure that no finite fraction of tile species dominates, with
tile fractions that are large compared to the remainder. Indeed, such
a tiling would essentially be a finite $D$ tiling, with a negligible
number of defect tiles added in.  Thus we introduce the notion of
``bounded'' relative fractions of tiles, which proves useful in
discussing large $D$ phason elastic constants. Consider tilings in
which all tile fractions tend to 0 as $D \ra \infty$, all vanishing at
the same rate. Specifically, assume there exists $a>0$ such that
\begin{equation}
n_{ij} \geq a \; n_{ij}^*
\label{cond2D}
\end{equation}
holds uniformly for all $i$ and $j$, which ensures that all tile
fractions have the same order of magnitude as in strain-free
tilings. Owing to eq. (\ref{nijdij}), we could alternatively assume
that there exists $b>0$ such that
\begin{equation}
d_{ij} \geq b \; d_{ij}^*.
\label{cond2Dbis}
\end{equation}
Indeed, $\sum_{k<l} n_{kl} = 1$ thus $\sum_{k<l} n_{kl} \sin |\pi(l-k)/D|
\leq 1$ and $d_{ij} \geq n_{ij}$. Furthermore $n_{ij} \geq a \; n_{ij}^*
= \pi a / 4 \; d_{ij}^*$ and $b = \pi a / 4$.

In addition, conditions~(\ref{cond2D}) or (\ref{cond2Dbis}) constrain
the distances $l_i$ and the angles $\theta_{ij}$ so that
\begin{equation}
l_i \leq c \; l_i^*
\label{encadre1}
\end{equation}
where 
$c>0$ is a finite constant, and
\begin{equation}
|\sin \theta_{ij}| \geq d \; |\sin \theta_{ij}^*|
\label{encadre2}
\end{equation}
with $d>0$. Indeed, these two conditions together with eq.
(\ref{dij}) imply $d_{ij} \geq (d/c^2) d^*_{ij}$.

Comparable constraints can be stated in the membrane viewpoint, in
which case the constraints appear as extreme allowed values of
$\vect{E}$.  Using (\ref{E_li_phii2}) and constraint~(\ref{encadre1}),
one finds that the phason strain is vanishingly small at large $D$,
\begin{equation}
\label{E2}
\vect{E}^2 \leq {Const. \over D}.
\end{equation}
The only way to achieve finite strain at large $D$ is to allow a
finite subset of the tile species to dominate. This large $D$ behavior
is related to our choice of normalizations of basis vectors both in
$\EC^\perp$ and in $\EC^\parallel$. Our choice is consistent with 
current practice for finite $D$ tilings.

\section{Mean-field theory, correlations and finite $D$ corrections}
\label{MFT}
\setcounter{equation}{0}

This section applies the iterative construction of
section~\ref{itere}, to develop a mean-field theory of strain-free
random tilings.  We first propose a close upper-bound of $\log 2$ for
the entropy per tile, which can be seen as the order 0 of the
mean-field theory developed below. We then refine the estimate by
calculating, within mean-field theory, the distribution of tile vertex
types. Finally we explore spatial correlations and related issues.

\subsection{First entropy estimate}
\label{order0}

Here we set the upper-bound of $\log 2$ for the entropy. To begin
with, we mention that rigorous bounds of the large $D$ entropy are
known in the case of {\em fixed-boundary tilings of unitary side
length} (see paper II~\cite{paperII}, section II):
\begin{equation}
\frac{1}{3} \log 2 \leq \sigma_{\infty}^{fixed} \leq
1.44 \log 2.
\end{equation}
Since it is demonstrated in the same reference that this entropy per tile
is equal to the free-boundary large $D$ entropy $\sigma_{\infty}$, the
same bounds should hold for $\sigma_{\infty}$. Our first mean-field
estimate $\log 2$ belongs to this interval, as well as the refined
mean-field estimate of the next section.

\begin{figure}[ht]
\begin{center}
\ \psfig{figure=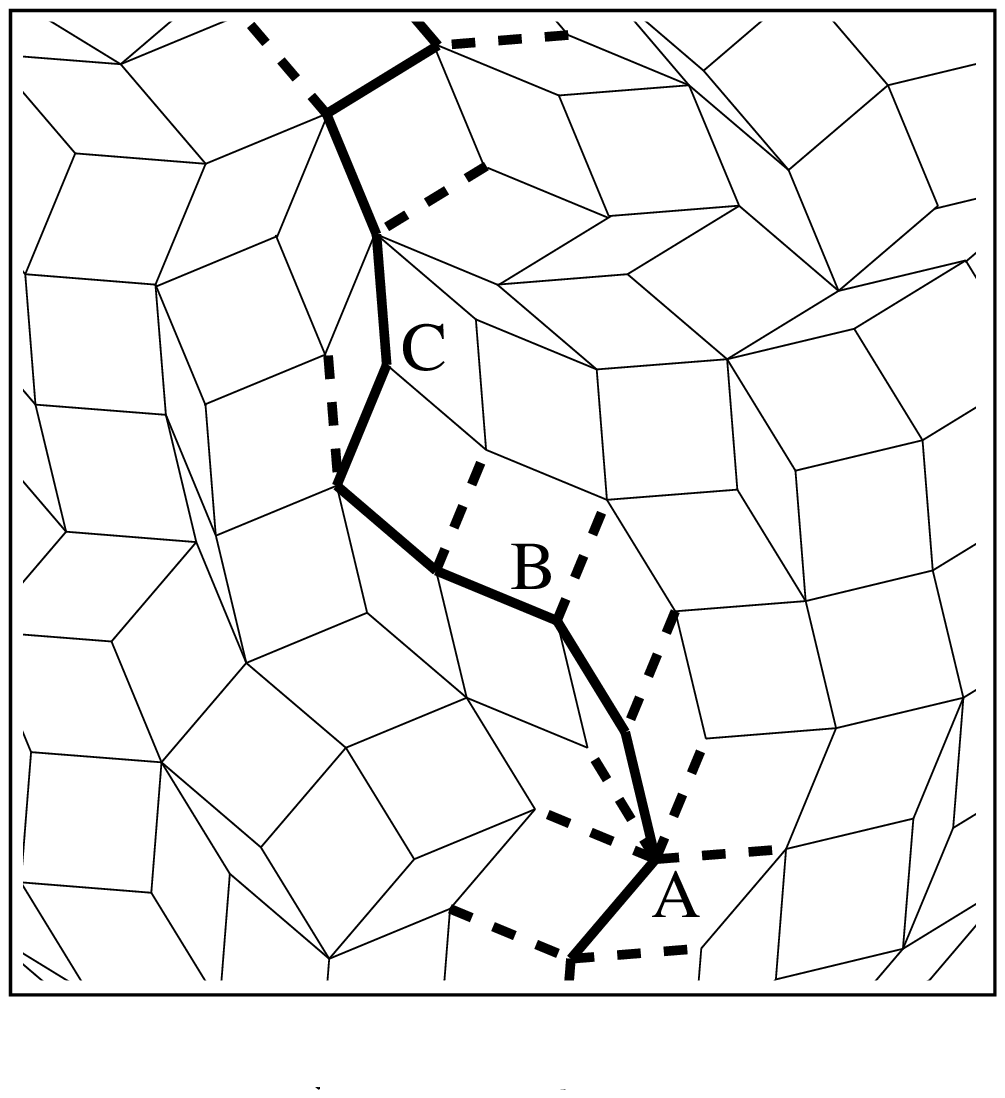,height=5cm} \
\end{center}
\caption{A random path (thick line) on a tiling as a succession of
vertices with choices (dashed edge = choice not taken). There are 5
choices at vertex A ({\em i.e.} $N_c(\mbox{A})=5$), 2 choices at vertex B and
only 1 choice at vertex C.}
\label{chemin}
\end{figure}

Let $P_D(N_D)$ denote the number of ways to follow a directed walk of
family $D$ on a $D \ra 2$ tiling. $N_D$ is the total length of a
single walk, or the cumulative length of a few walks. We will neglect
contacts among walks of family $D$, which is justified since $l_i \sim
D$.  To calculate a typical value of $P_D(N_D)$, consider the problem
of constructing bottom-to-top paths on a typical tiling. At each step,
the path may follow one or more routes (Fig.~\ref{chemin}). The only
requirement is that the path segment leading out of a vertex must
contain an upwards component. We estimate the number of paths as the
product over vertices $v$ of the number of choices $N_c(v)$ to be made
at each step
\begin{equation}
\label{eq:gm}
P_D(N_D) \approx \prod_{v=1}^{N_D} N_c(v).
\end{equation}

Evaluation of equation~(\ref{eq:gm}) requires the distribution of
values $N_c(v)$ along paths. Because we do not know this distribution
(yet), we settle for an estimate that yields an upper bound on
$P_D(N_D)$, and eventually on the entropy per tile $S_D$.  Note that
the product in equation~(\ref{eq:gm}) is the $N_D$-th power of the
geometric mean of $N_c(v)$. The geometric mean of any set of positive
numbers is bounded above by the arithmetic mean, reaching this bound
only when all values are equal.  The arithmetic mean
$\bar{N_c}=\bar{Z}/2$, with $\bar{Z}$ the mean coordination number (or
degree), because on average half the tile edges at each vertex have a
vertical component in the northerly direction.  From Euler's theorem
applied to infinite rhombus tilings, we know $\bar{Z} = 4$, so that
$\bar{N_c} = 2$.  We deduce the $D$-independent upper bound
\begin{equation}
\label{pathbound}
P_D(N_D) \le 2^{N_D}.
\end{equation}

Let $B_D$ be the number of $D \ra 2$ tilings (free boundary tilings of
some size $\RC$).  Since each $D+1 \ra 2$ tiling is in one-to-one
correspondence with random walks on a $D \ra 2$ tiling,
\begin{equation}
B_{D+1} \approx P_D(N_D) B_D.
\label{B_ratio}
\end{equation}
We iterate this relationship to write
\begin{equation}
B_D=\prod_{D'=3}^{D-1} P_{D'}(N_{D'}).
\label{B_D}
\end{equation}
Inserting relation~(\ref{pathbound}) for the path counting, we find
\begin{equation}
B_D \leq 2^{N_3+\ldots +N_{D-1}}.
\end{equation}
Since the number of tiles is $N=\sum_{D'} N_{D'}$, we obtain the
entropy per tile
\begin{equation}
\sigma_D = {{\log{B_D}}\over{\sum N_{D'}}} \le \log{2} = 0.693.
\label{eq:our_bound}
\end{equation}
The limiting entropy $\sigma_{\infty}$ is bounded above by $\log 2$.

\subsection{Refined mean-field calculations}
\label{MFcalc}

The entropy value $\log{2}$ just derived is not exact for two
reasons. First, to properly compute the geometric mean of $N_c$ we
need the full vertex path choice probability distribution
$\pi(N_c)$. Second, our estimate eq.~(\ref{eq:gm}) neglected
correlations among the numbers of choices at different steps along the
path. In this subsection we address the first point. In the next one
we shall address briefly the second point.

Following the logic in section~\ref{order0}, the limiting entropy is
\begin{equation}
\sigma_\infty = \lim_{D \ra \infty} \lim_{k \ra \infty} {\log P_D(k) \over k},
\label{sigma1}
\end{equation}
where $P_D(k)$ is the number of $k$-step paths on a $D \ra 2$ tiling.
The mean-field approximation assumes that the steps of such paths are
uncorrelated so that the number of paths is given by
eq.~(\ref{eq:gm}).  The vertices $v$ belong to a random $D \ra 2$
tiling. Therefore the numbers of choices $N_c(v)$ are distributed
according to a probability distribution $\pi_D(N_c)$. When $D$ tends
to infinity, this distribution tends toward a limiting distribution,
denoted by $\pi(N_c)$. Thus, after a short calculation, equations
(\ref{sigma1}) and (\ref{eq:gm}) become:
\begin{equation}
\sigma_\infty^{MF} = \sum_{N_c=1}^{\infty} \pi(N_c) \log N_c. 
\label{sigma2}
\end{equation}

\begin{figure}[ht]
\begin{center}
\ \psfig{figure=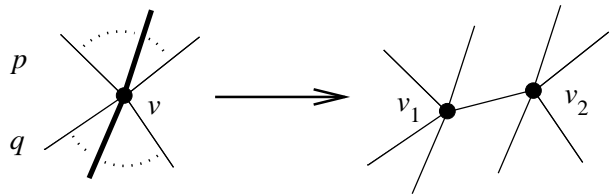,width=8cm} \
\end{center}
\caption{A path going through a vertex $v$ of a $D \ra 2$ tiling 
gives birth to two vertices $v_1$ and $v_2$ of a $D+1 \ra 2$ tiling.
The path follows any leg of $v$ with probability $1/q$ 
and any arm with probability $1/p$.}
\label{meanfield}
\end{figure}

To get the mean-field distribution $\pi(N_c)$, let us first denote by
$\pi_D(q,p)$ the fraction of vertices on a $D \ra 2$ random tiling
with $q$ in-coming edges (``legs'') and $p$ out-going ones (``arms'')
(see figure~\ref{meanfield}).  Neglecting correlations, a
$(q,p)$-vertex $v$ of a $D \ra 2$ tiling will be visited by paths with
probability $\pi_D(q,p)$ and will then give birth to two vertices
$v_1$ and $v_2$, as illustrated in figure~\ref{meanfield}. These
latter vertices belong to a $D+1 \ra 2$ tiling. Each leg and each arm
of $v$ will be chosen with probabilities $1/q$ and $1/p$,
respectively. The new edge from $v_1$ to $v_2$ is arbitrarily chosen
to be oriented upward in the $D+1 \ra 2$ tiling and is therefore an
arm for $v_1$ and a leg for $v_2$.

Indeed, a vertex with $p'$ arms and $q'$ legs can be either the 
``left-son'' $v_1$ or the ``right-son'' $v_2$ of a ``father-vertex''
$v$ in a $D \ra 2$ tiling (figure~\ref{meanfield}). 
The two son vertices together inherit all their father's
arms and legs, plus an extra arm and leg (for the path along which the
vertex is split).  Finally, in addition, the left son gets an
extra arm and the right son an extra leg from the new bond that joins them.
Consequently the left-son has more than one arm and the right-son has
more than one leg.

As a consequence, in this mean-field approximation, the probabilities
$\pi_{D+1}(q,p)$ can be written as linear combinations of the
probabilities $\pi_D(q,p)$. The corresponding linear operator will be
denoted by $\AC: \pi_D \mapsto \pi_{D+1}$. It is infinite-dimensional
since $q$ and $p$ can be arbitrarily large when $D$ goes to
infinity. The operator $\AC$ is defined by:
\begin{equation}
\pi_{D+1}(q',p') = (1-\delta_{q',1}) \sum_{q \geq q'-1 \atop p \geq
p'} {\pi_D(q,p) \over 2 q p} \ +(1-\delta_{p',1}) \sum_{p \geq p'-1
\atop q \geq q'} {\pi_D(q,p) \over 2 q p}.
\label{operator:A}
\end{equation}
The first and second term refer, respectively, to left- and right-sons
($v_1$ and $v_2$). The factors of $1/2$ expresses that a vertex is either a
left-son or a right-son with probability $1/2$. Moreover, each arm-leg
pair is chosen with probability $1/pq$ among all possibilities
(see figure~\ref{meanfield}).  The
limiting distribution $\pi(q,p)$ is the fixed point of $\AC$, the
eigenvector with eigenvalue 1.

The diagonalization of $\AC$ being rather complex, we focus on a
summation of the previous relation on legs $q$. This yields a
recursion relation on the arm statistics, in other words on the choice
statistics
\begin{equation}
\pi_D(N_c)= \sum_{q=1}^{\infty} \pi_D(q,N_c).
\end{equation}
We denote this operator by $\hat{\AC}$ and write it in matrix form,
where rows and columns are indexed by $N_c$, as
\begin{equation}
\label{AC}
\hat{\AC} = \left(
\begin{array}{cccccc}
{1 \over 2} 1 &{1 \over 2}{1 \over 2} &{1 \over 2}{1 \over 3} 
&{1 \over 2}{1 \over 4} & {1 \over 2}{1 \over 5} &\cdots \vspace{1mm} \\
{1 \over 2} 1 &{1 \over 2} &{1 \over 3} &{1 \over 4} & {1 \over 5} 
&\cdots \vspace{1mm} \\
0&{1 \over 2}{1 \over 2} &{1 \over 3} &{1 \over 4} & {1 \over 5} &\cdots 
\vspace{1mm} \\
0& 0 &{1 \over 2}{1 \over 3} &{1 \over 4} & {1 \over 5} &\cdots \vspace{1mm} \\
0 & 0 & 0 &{1 \over 2}{1 \over 4} & {1 \over 5} & \cdots \vspace{1mm} \\ 
0 & 0 &0 &0 & {1 \over 2}{1 \over 5}& \cdots \\
\vdots & \vdots &\vdots & \vdots& & \ddots \\
\end{array}
\right).
\end{equation} 
This operator preserves the mean number of choices $\langle N_c\rangle=2$.  The
fixed point is
\begin{equation}
\pi(N_c) = {1 \over \sqrt{e}} \; {2 N_c -1 \over 2^{N_c} (N_c -1)!}.
\label{distrib:Nc}
\end{equation}
The first few values of $\pi(N_c)$ are listed in table~\ref{results},
where they may be compared with Monte Carlo simulation results (see
paper II~\cite{paperII}). The values peak at $N_c=2$ and decrease
rapidly for large $N_c$.

The mean-field entropy in eq.~(\ref{sigma2}) becomes
\begin{equation}
\sigma_\infty^{MF} =
{1 \over \sqrt{e}} \; \sum_{N_c=1}^{\infty} 
{2 N_c -1 \over 2^{N_c} (N_c -1)!} \log N_c \simeq 0.598.
\label{ent_mft}
\end{equation}
This mean-field value is satisfactorily close to the numerical value
obtained by Monte Carlo simulations (see paper II~\cite{paperII}),
namely $\sinf \simeq 0.568$, and is well below the upper bound $\log{2}=0.693$.

\begin{table}[htb]
\caption{The first values of the limiting distribution $\pi(N_c)$, obtained 
both in the mean-field approximation eq.~(\ref{distrib:Nc}) and
numerically, by Monte Carlo simulations (from paper
II~\cite{paperII}).}
\begin{center}
\begin{tabular}{|l|r|r|r|r|r|}
\hline
$N_c$                   & 1 & 2 & 3 & 4 & 5 \\
\hline
$\pi(N_c)$ (Mean-field) & 0.30 & 0.45 & 0.19 & 0.04 & 0.007\\
\hline 
$\pi(N_c)$ (Numerical)  & 0.26 & 0.51 & 0.21 & 0.03 & 0.001 \\
\hline
\end{tabular} 
\end{center}
\label{results}
\end{table}

\subsection{Spatial correlations}
\label{correl}

Spatial correlations play a crucial role in two-dimensional finite
codimension tilings since they are critical systems with long-range
correlations~\cite{blote}. In particular, they condition the diffusive
behavior of de Bruijn lines, they constrain the successive steps in
the iterative construction process, and they reduce the entropy
slightly below the mean-field value. We can calculate both short- and
long-range correlations among $p$ and $q$ within our mean-field
theory.  We also note some expected properties of perpendicular space
height correlations.

\subsubsection{Short-ranged correlations}
\label{short-range}

Consider the short-ranged correlation between the numbers of arms and
legs $p$ and $q$ on individual vertices.  Within the mean-field
approximation, $\langle pq\rangle_D = \sum_{p,q}\pi_D(q,p) pq$. If, at
step $D$, a vertex $v$ has $p$ arms and $q$ legs (see
Fig.~\ref{meanfield}), then its first son $v_1$ has on average $(2 + 3
+ \ldots (p+1))/p=(p+3)/2$ arms independently of its number of legs,
and $(1+\ldots+q)/q=(q+1)/2$ legs independently of its number of
arms. Similarly, its second son $v_2$ has on average $(p+1)/2$ arms
and $(q+3)/2$ legs. Therefore
\begin{eqnarray}
\langle pq\rangle_{D+1} & = & \sum_{p,q} \pi_D(q,p) \left( 
{1 \over 2} \; {p + 1 \over 2} \; {q + 3 \over 2} +
{1 \over 2} \; {p + 3 \over 2} \; {q + 1 \over 2} \right) \\ \nonumber
 & = & {1 \over 4} ( \langle pq\rangle_D + 11),
\end{eqnarray}
since $\langle p\rangle _D=\langle q\rangle _D=\langle N_c\rangle
_D=2$.  The limiting value is $\langle pq\rangle _{\infty} =
11/3$. Moreover, distribution~(\ref{distrib:Nc}) leads to $\Delta p =
(\langle p^2\rangle _{\infty}-\langle p\rangle
_{\infty}^2)^{1/2}=\sqrt{3}/2$ and $\Delta p =\Delta q$. Therefore the
mean-field covariance of $p$'s and $q$'s is
\begin{equation}
\cov(p,q) = { \langle pq\rangle _{\infty} - \langle p\rangle _{\infty} 
\langle q\rangle_{\infty} \over \Delta p \, \Delta q} = -4/9 \simeq -0.44, 
\end{equation}
also in good agreement with the numerical value $\cov(p,q) \simeq
-0.36$ (see paper II~\cite{paperII}). The main contribution to those
correlations comes from the asymmetry caused by the edge between $v_1$
and $v_2$.

We can use the covariance to improve on our original ``upper bound''
entropy estimate eq.~(\ref{eq:our_bound}) of $\log{2}=0.693$.  Whereas
the original estimate arose from the mean number of choices in single
steps, we can instead consider the geometric mean number of ways to
make two steps.  The product $pq$ counts the number of two-step paths
passing through a given vertex. Hence an improved entropy estimate is
$\log{\sqrt{11/3}}=0.649$, which is indeed closer to the numerical
value of 0.568, though nowhere near as close as our full mean-field
value in eq.~(\ref{ent_mft}).  Ideally, we would determine the full
distribution of choices (i.e. number of arms) in two-step paths,
$\pi_2(N_c^1,N_c^2)$ then we could replace our MFT estimate
eq.~(\ref{ent_mft}) with
\begin{equation}
\label{ent_2step}
\sigma_{\infty}^{MF2}=\sum_{N_c^1,N_c^2=1}^{\infty}
\pi_2(N_c^1,N_c^2) \log\sqrt{N_c^1 N_c^2},
\end{equation}
which should improve upon the value in eq.~(\ref{ent_mft}) just
as $\log{\sqrt{11/3}}$ improved upon $\log{2}$.  Unfortunately,
it is considerably more difficult to obtain $\pi_2(N_c^1,N_c^2)$
than $\pi(N_c)$.

\subsubsection{Long-ranged correlations}
\label{long-range}

To calculate long-range correlations using our iterative construction
method, two phenomena compete: de Bruijn lines running through a
tiling destroy vertex-to-vertex correlations, while they increase
distances between vertices.  Consider two vertices separated by a
distance $u_n$ on an $n \ra 2$ tiling.  The addition of successive
families of lines to get higher codimension tilings ($D \ra 2$ with
$D>n$) make the tiling ``swell'' homogeneously, and the distances grow
like $D$.  When $D$ increases, the distance between the two vertices
increases so that
\begin{equation}
\label{scale}
u_D / D \simeq u_n / n.
\end{equation}
The idea is now to track such pairs of vertices, and to estimate how
their correlation functions evolve at the same time they are moving
apart.

To track these correlation functions as $D$ increases, we estimate
correlations between the number of choices at vertices, $N_c(v)$.
Define the correlation function between vertices $v_a$ and $v_b$,
widely separated in space,
\begin{equation}
C_n(r)={1 \over N_P} \sum_{d(v_a,v_b)=r} N_c(v_a) N_c(v_b) -
\left({1 \over N_v} \sum_v N_c(v) \right)^2 
= \langle N_c(v_a) N_c(v_b) \rangle_r  - \langle N_c\rangle^2.
\end{equation}
The first average runs over the $N_P$ pairs of vertices of $n \ra 2$
tilings separated by a distance $r$.  The second average runs over all
$N_v$ vertices in the tiling.  In the following, we denote $\langle N_c(v_a)
N_c(v_b) \rangle_r = \gamma_n$.  Moreover the mean number of choices
$\langle N_c\rangle=2$, so that
\begin{equation}
C_n(r) = \gamma_n - 4.
\end{equation}

When a new line passes through either $v_a$ or $v_b$, their
correlation diminishes. Each vertex has a probability $2/n$ that a new
line goes through it (see eq.~(\ref{meanni})). Therefore the
probability that either one of the pair $(v_a,v_b)$ will be affected
by a new line is $4/n$. A fraction $1-4/n$ of pairs separated by $r$
are unaffected, so their correlation remains equal to $\gamma_n$. The
remaining pairs {\em are} affected, and we denote their new
correlation by $\gamma'_n$ which we now calculate.

Each pair $(v_a,v_b)$ that is affected gives birth to two pairs.  One
of its vertices, say $v_b$, remains unchanged while the other one,
$v_a$, gives birth to two vertices, a ``left-son'' $v_{a1}$ and a
``right-son'' $v_{a2}$ (see Fig.~\ref{meanfield}). On average,
\begin{equation}
N_c(v_{a1}) = { 1 + 2 + \ldots + N_c(v_a) \over N_c(v_a)} = N_c(v_a)/2 + 1/2,
\end{equation}
and $N_c(v_{a2}) = N_c(v_a)/2 + 3/2$, since a right-son has one more
arm than a left-son.  Therefore, averaging on the whole tiling, we find
\begin{equation}
\gamma'_n = 1/2 ( \langle N_c(v_{a1}) N_c(v_b)\rangle + \langle N_c(v_{a2}) N_c(v_b)\rangle ) =
\gamma_n/2 +2.
\end{equation}

Now we estimate $\gamma_{n+1}$, the correlation averaged over all
tiling vertices after the introduction of the lines of family $n+1$.
For original vertices {\em not} touched by the new lines, the
correlation remains $\gamma_n$. For original vertices that {\em are}
touched by the new lines, the correlation becomes $\gamma'_n$.
Therefore,
\begin{equation}
\gamma_{n+1}={(1 - 4/n) \gamma_n + 8/n \; \gamma'_n \over (1 - 4/n) + 8/n}
\approx (1 - 8/n) \gamma_n + 8/n \; \gamma'_n + \OC(1/n).
\end{equation}
Hence
\begin{equation}
C_{n+1}(u_{n+1}) = (1 - 4/n) C_n(u_n).
\end{equation}
Now iterating from $n \ra D$ we write
\begin{equation}
C_D(u_D) = C_n(u_n) \prod_{k=n+1}^D (1 - 4/k).
\end{equation}
For large $n$ and $D$,
\begin{equation}
\label{logCD}
\log C_D = \log C_n + \sum_{k=n}^{D-1} \log (1 - 4/k)
\simeq \log C_n - 4 \sum_{k=n}^{D-1} 1/k
\simeq \log C_n - 4 \log (D/n)
\end{equation}
from which it follows that 
\begin{equation}
\label{D4CD}
D^4 \; C_D(u_D) \simeq n^4 \; C_n(u_n).
\end{equation}
Using eq.~(\ref{scale}) we have
\begin{equation}
(u_D)^4 \; C_D(u_D) \simeq (u_n)^4 \; C_n(u_n).
\end{equation}
and finally
\begin{equation}
C_{\infty}(r) \simeq const. \; {1 \over r^4}.
\end{equation}
This expression holds only for large values of $r$, since we do not
take into account the discreteness of tilings at very short range.
Correlations fall off rapidly at large distance.

\subsubsection{Height-height correlations}
\label{height-height}

In two dimensions, the random tiling hypothesis eq.~(\ref{hypothesis})
imples logarithmic behavior of the two-point height-height correlation
function~\cite{Henley91}
\begin{equation}
G(r) \equiv \langle(h(r)-h(0)^2\rangle = {{1}\over{\pi K}} \log{r}
\label{eq:height-height}
\end{equation}
in the large $r$ limit.  We give here the generic form, without regard
to the specific components of our $D-2$ dimensional height function or
the particular irreducible components of the phason strain tensor
${\bf K}$.  The only requirement is that the elasticity tensor be
positive definite, i.e. $K>0$.  Presumably
eq.~(\ref{eq:height-height}) will hold for any finite $D$.

We can easily derive an alternative formula for $G(r)$ in the limit $r \ll D$.
%NICOLAS: How does $r$ in the tiling space transform into the grid space,
%where I am really doing the calculation? Also please help check the
%scaling of |h|^2.
Consider the shortest path joing two tiling vertices, and let $s$ be
the number of tile edges on this path.  Provided $s \ll D$, a typical
path contains $s$ distinct edge orientations $\pm \eh_i^\parallel$ and
its length $r$ is proportional to $s$. In the perpendicular space,
$h(s)-h(0)$ contains $s$ distinct terms equaling $\pm \eh_i^\perp$, the
remaining $D-2-s$ components equaling 0. Since the products 
(\ref{double}) are of order at most $1/D$ whereas $(\eh_i^\perp)^2=1$,
$(h(s)-h(0))^2 \simeq s$ when $s \ll D$. Consequently
\begin{equation}
G(r) \propto r
\label{eq:small_r}
\end{equation}
for $r\ll D$.  Now taking the $D\ra\infty$ limit, and comparing
eq.~(\ref{eq:small_r}) with the large $r$
limit~(\ref{eq:height-height}), we see that $K$ must vanish in the
limit of large $D$ or else $G(r)$ will lack monotonicity.
Section~\ref{elastic:tensor} provides further evidence for the
vanishing of $K$.

\subsection{Finite $D$ corrections}
\label{finiteD}

The mean-field calculations of section~\ref{MFcalc} neglect the fact
that at the step $D$ of the iterative process, only a diminishing fraction of
of the vertices are visited by paths. Correcting this
oversight does not alter the fixed point of the process, but predicts
finite $D$ corrections to it. Indeed, at step $D$, a fraction $2/D$ of
the vertices are visited. Each of them has 2 sons and
equation~(\ref{operator:A}) applies only to a fraction of new vertices of
order $4/D$ when one iterates at step $D$. The remaining vertices
are unaffected. Hence the correct evolution operator is
\begin{equation}
\label{hat:A}
\hat{\AC}_D = {4 \over D} \hat{\AC} + \left( 1 - {4 \over D} \right) \mbox{Id}.
\end{equation}
where $\hat{\AC}$ is the operator introduced in eq.~(\ref{AC}). For any
finite $D$, the fixed point of $\hat{\AC}$ remains the fixed point of
$\hat{\AC}_D$. Any eigenvector $e$ of $\hat{\AC}$ associated with the
eigenvalue $\mu$ evolves as follows:
\begin{equation}
\label{prod:hat:A}
\hat{\AC}_D \hat{\AC}_{D-1} \ldots \hat{\AC}_2 \cdot e  =   
\prod_{k=2}^D \left( {4 \over k}\mu + ( 1 - {4 \over k}) \right) 
e \ =   
\prod_{k=2}^D \left( 1 + {4(\mu -1) \over k} \right) e 
\ \simeq {C \over D^{4(1-\mu)}} \; e,
\end{equation}
where $C$ is some constant of order 1, using the same calculation
scheme as in equations (\ref{logCD}) and (\ref{D4CD}).

Hence we get power-law convergence towards the fixed point instead of
exponential convergence. The spectrum of $\hat{\AC}$ can be calculated
numerically, by taking finite but larger and larger approximations of
$\AC$. The spectrum converges rapidly towards $\{
1,1/2,1/3,\ldots,1/p,\ldots\}$.  Eigenvalue $\mu=1$ is the desired
fixed point.  The slowest mode corresponds to $\mu = 1/2$ and to a
$1/D^2$ behavior of the finite $D$ corrections to $\pi(N_c)$, and
therefore to the entropy~(\ref{sigma2}). Thus finite $D$ corrections
to the entropy associated with path-counting are expected to behave
like $1/D^2$ in the mean-field approximation.

\section{de Bruijn line interactions}
\label{interactions}
\setcounter{equation}{0}

The mean-field theory addresses the statistics of isolated de Bruijn
lines influenced by the statistics of the $D-1\ra 2$ tiling they
meander on, but with no other interactions.  A crucial fact about the
de Bruijn line construction of tilings is that lines within the same
family may never cross, resulting in an effective repulsive
interaction at contact~\cite{Fisher84}.  We now analyze the effect of
interactions on finite size corrections and on phason elasticity.

\subsection{Entropy loss due to contacts in strain-free tilings}
\label{contacts:metric}

We call ``contact'' between two neighbor lines of a same family a
configuration where these lines are adjacent. Such a situation is
represented in figure~\ref{contact}, where we have circled one contact
between neighboring paths (or de Bruijn lines). If we follow the lines
in the direction $t$, the local initial angles $\theta$ and $\theta'$
they make with their global direction $\vect{\delta}$ after the
contact are constrained by the noncrossing relation $\theta \leq
\theta'$. This constraint divides the number of allowed configurations
by a factor of order 2 as compared to the free path case, and reduces
the total entropy by about $\log 2$.

To estimate the global entropy loss due to contacts in an infinite
tiling, we must calculate the density of contacts. A crucial quantity
is the distance $L_i$ between consecutive contacts with neighboring
lines of family $i$. Inspecting Fig.~\ref{contact} we see that
contacts between lines of family $i=D$ result from diffusion of the
lines in the direction $\eh^\parallel_D$. Note that $L_i$ depends on
both $l_i$ (because wider separation reduces the frequency of contacts
of diffusing lines) and $\phi_i$ (because the diffusion constant has
corrections in the $\phi_i$; these corrections are quadratic because
of the symmetry of the problem through any line).
%NICOLAS: I can't find that explanation anywhere. I want to give
%very short explanation here.
In strain-free tilings, which have maximal rotational symmetry,
$l_i=l_i^*=D/2$ and $\phi_i=0$ and $L_i=L$ is independent of $i$.

Define the diffusion constant ${\cal D}$ so that
the lateral displacement $u$ after traveling a distance $L$ in the
direction $\delta_i$ is $u^2 \sim {\cal D} L$. We presume that ${\cal
D}$ takes a finite limiting value in the large $D$ limit. Thus ${\cal
D} L \sim (l_i^*)^2$, and there is approximately one contact
among lines of family $i$ in every region of area ${\cal A} = l_i^* L
= (l_i^*)^3/{\cal D}$.  Adding up the total number of contacts
expected among lines of all orientations, we estimate the loss of
entropy per unit area as
\begin{equation}
\label{Delta:s:E0}
\Delta \sigma \approx 
- {{\cal D} D \over (l_i^*)^3} \log 2 = - 8 {\cal D} \log 2 {1 \over D^2}.
\end{equation}
This quantity should be suitably rescaled if one is interested in the
entropy per tile. The strain-free loss of entropy falls off like
$1/D^2$ because as $D$ grows large, lines of like orientation rarely
contact each other.

\subsection{Phason elastic constant tensor}
\label{elastic:tensor}

We now investigate the phason strain dependence of the entropy and
test the generic form assumed in eq.~(\ref{hypothesis}).  A thorough
discussion of phason elasticity requires a group-theoretical analysis
of rotational invariants~\cite{Socolar85}, each leading to an
independent phason elastic constant, which lies beyond the scope of
this paper.  Rather, we will assume a generic form of strain and
estimate its impact on the entropy.  In particular, we shall see that
${\bf K}$ decays like $1/D$ when $D$ becomes large.

To start, consider expanding $\sigma_D(\vect{E})$ to second order in
$\phi_i$ and $\delta l_i=l_i -l_i^*$.  Note that this expansion does
not couple explicitly angles $\phi_i$ and line spacings $\delta l_i$
because a symmetry through any line changes $\phi_i$ into $-\phi_i$
whereas it does not affect $\delta l_i$.  Therefore this expansion is
the sum of a quadratic form in the $\phi_i$'s and of a quadratic form
in the $\delta l_i$'s.  Some phason elastic constants are related to
the angles $\phi_i$ only.  The remaining ones are related to the
variations of the line spacings $l_i$ only.

At present we consider only the $l_i$ strain.  To estimate the
strain-dependence of the loss of entropy by contact we must calculate
the density of contacts as a function of strain. We
impose a global phason strain by varying only the mean distance $l_i$
between de Bruijn lines of family $i$.  Let
\begin{equation}
\label{smallstrain}
l_i=l_i^*(1+\lambda \delta_i)
\end{equation}
where $\lambda$ is a small parameter that controls the strength of the
phason strain. Here $\delta_i$ is an arbitrary but bounded function of
$i$ whose mean and standard deviation obey $\langle \delta\rangle=0$
and $\langle \delta^2\rangle=1$, where the averages are taken over
families $i$.  Owing to the bounded tile fraction constraint
eq.~(\ref{encadre1}), whatever the strength of the phason strain,
$l_i$ is of order $D$ and $\lambda$ is of order at most 1.  The
requirement of vanishing mean is imposed by the normalization
condition (\ref{normalize}).

Consider a pair of adjacent lines of family $i$, and evaluate $L_i$,
the mean distance between contacts of these lines.  Thus ${\cal D} L_i
\sim l_i^2$, and there is approximately one contact among lines of
family $i$ in every region of area ${\cal A}_i = l_i L_i = l_i^3/{\cal
D}$.  The entropy loss (\ref{Delta:s:E0}) becomes
\begin{equation}
\label{pstrain}
\Delta \sigma \approx 
- {\cal D} D \langle l^{-3}\rangle \log 2.
\end{equation}
Now consider the average $\langle l^{-3}\rangle$.  For vanishing phason strain
($\lambda \ra 0$), eq.~(\ref{li}) reveals that $l_i=l_i^*=D/2$.
Expanding $l_i^{-3}$ in eq.~(\ref{smallstrain}) for small $\lambda$,
using the known mean and standard deviation of $\delta_i$, we obtain
\begin{equation}
\label{liaverage}
\langle l^{-3}\rangle \approx \left( {{2}\over{D}} \right)^3 (1+6 \lambda^2).
\end{equation}
Substituting this expression into eq.~(\ref{pstrain}) yields
\begin{equation}
\label{elastic}
\Delta \sigma \approx 
- \left( {{1}\over{D^2}}+{{6 \lambda^2}\over{D^2}} \right)
8 {\cal D} \log{2}.
\end{equation}
We find two terms, each falling off as $1/D^2$. The first term, which
arises in the absence of strain, reproduces eq.~(\ref{Delta:s:E0}).
In a strained tiling with $\lambda \ne 0$, owing to
eq.~(\ref{encadre1}), $l_i$ remains of order $D$, and the loss of
entropy is also of order $1/D^2$.  Although the frequency of contacts
increases due to the strain, it still falls off for large $D$.

Because the entropy is independent of the phason strain at the large
$D$ limit, we again conclude that the phason elastic constant
vanishes.  We have seen (eq.~(\ref{E2})) that for bounded tile
fractions the phason strain $\vect{E}^2$ decreases like $1/D$ as $D$
increases.  Comparing eq.~(\ref{hypothesis}) with the
strain-corrections of order $1/D^2$ that we can calculate, we conclude
that the phason elastic tensor ${\bf K}$ falls off as $1/D$.

We can understand the vanishing phason strain correction by the
following qualitative argument.  Consider a domain of the tiling of
diameter $\Delta$, large as compared to the tile size 1 but small as
compared to the mean distance between two lines of a same family, $1
\ll \Delta \ll D$.  For example, one can choose $\Delta = \sqrt{D}$.
In such a domain there is {\em at most} one de Bruijn line per family,
and no lines at all of most families. Since $\Delta$ tends to infinity
with $D$, the area of the domain tends to infinity and the number of
lines and tiles in the domain also tend to infinity.

\begin{figure}[ht]
\begin{center}
\ \psfig{figure=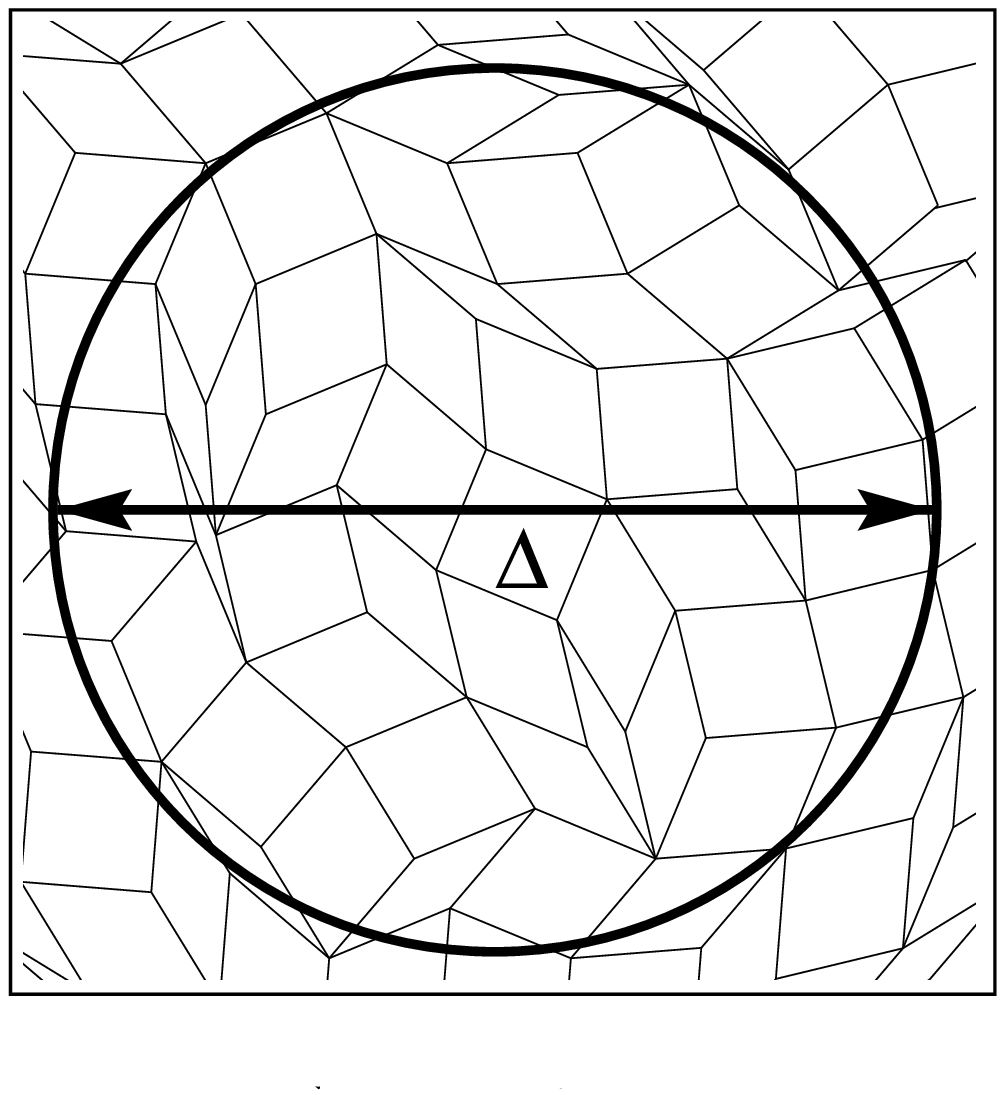,width=5cm} \
\end{center}
\caption{A domain of diameter $\Delta$ which contains at most one de
  Bruijn line per family.}
\label{Delta}
\end{figure}

Locally, as $D$ and $\Delta$ go to infinity, the tiling looks like a
large codimension $D' \ra 2$ tiling (where $D' \sim \Delta$), with
only one line per family (see Fig.~\ref{Delta}).  The entropy per tile
in this domain tends towards a quantity $\sigma_{loc}$ which depends
neither on the strain nor on the domain under consideration and is
directly related to a simplified problem with one line in each family.
The global entropy per tile $\sinf$ is the average over the whole
tiling~\cite{Bibi97} of the local entropy density $\sigma_{loc}$.
Therefore $\sinf = \sigma_{loc}$ is independent of the de Bruijn line
separations $l_i$, of the angles $\phi_i$ and of the global phason
strain.

\section{Tilings of dimension other than 2}
\label{dnot2}
\setcounter{equation}{0}

This section discusses briefly cases of dimension other than
$d=2$. The high codimension $d=1$ case is especially interesting
because it is exactly solvable. It serves as an excellent illustration
of several of the concepts in the paper and contrasts interestingly
with the high codimension $d=2$ case. We then present results and
conjectures for the high codimension $d=3$ case and the codimension 1
case with $d=D-1$ for arbitrary $d$.

\subsection{$D \ra 1$ tilings}
\label{1d}

One-dimensional high codimension $D \ra 1$ tilings are exactly
solvable, as we show it in this subsection.  The special case $D=2$
was previously discussed, for example in~\cite{Henley91}.  These
tilings comprise $D$ different species of tiles randomly arranged on a
line. This case is unusual since the entropy per tile tends to
infinity with $D$. Still, this example is quite instructive since it
already contains a characteristic feature of large codimension
tilings: when $D$ goes to infinity, we prove that the entropy no
longer depends on the relative tile fractions provided these
quantities remain bounded.

The number of tilings is given by the multinomial coefficient:
\begin{equation}
W^{D \ra 1}(k_1,\ldots,k_D) = 
{(k_1 + k_2 + \ldots + k_D)!  \over k_1! \; k_2 ! \ldots k_D !}.
\end{equation}
Introducing tile fractions $x_i = k_i/(k_1+\ldots+k_D)$, 
we get the entropy per tile at large $D$
\begin{equation}
\sigma^{D \ra 1}=-\sum_{i=1}^D x_i \log x_i.
\end{equation}
This entropy is maximal at $x_i=1/D$ where it equals $\log D$.  In one
dimension, for the sake of convenience in the calculation, we set the
upper bound $x_i \leq a /D$ where $a > 0$ to prevent some tile species
from being dominant.  Then $- \log x_i \geq \log D - \log a$, and
hence $-\sum_{i=1}^D x_i \log x_i \geq \log D - \log a$ because
$\sum_{i=1}^D x_i = 1$. Therefore
\begin{equation}
\log D - \log a \leq \sigma^{D \ra 1} \leq \log D.
\end{equation}
The entropy grows asymptotically like $\log D + \OC(1)$, independently of
the exact tile fractions provided they remain bounded.

Large $D$ phason elastic constants can also be calculated. Near the
entropy maximum at $x_i=1/D$,
\begin{equation}
K_i = {\partial^2 \sigma \over \partial x_i^2} = -D. 
\end{equation}
Let $x_i={{1}\over{D}}+\delta x_i$, so that $\sigma=\maxindex{\sigma}
- {{D}\over{2}} \sum \delta x_i^2$. The average gradient of the function $h$
(representing the one-dimensional membrane in the $D$-dimensional
space) is characterized by the average angle $\psi$ this membrane
makes with the reference direction $(1,1,\ldots,1)$. Specifically,
$(\nabla h)^2 = \tan^2 \psi \simeq \psi^2$. Now,
\begin{equation}
1 - {\psi^2 \over 2} \simeq \cos \psi = \left(D \sum_{i=1}^D x_i^2
\right)^{-1/2} \simeq 1 - {D \over 2} \sum_{i=1}^D \delta x_i^2.
\end{equation}
Finally, $(\nabla h)^2 = D \sum \delta x_i^2$ and
\begin{equation}
\sigma^{D \ra 1} \simeq \maxindex{\sigma} - {K \over 2} (\nabla h)^2,
\end{equation}
where the phason elastic constant $K=1$ is vanishingly small as
compared to the entropy $\sigma_{\mbox{\scriptsize{max}}}$.  Note the
value of $K$ is independent of $D$, and of course it agrees with the
previous value~\cite{Henley91} derived for $D=2$.

\subsection{$D \ra 3$ tilings}
\label{D3tilings}

Unlike the cases of $d=1$ and $d=2$, we have neither exact, nor
extensive numerical information available for high codimension $d=3$
tilings. Table~\ref{tab:d3} collects available data on $d=3$
strain-free boundary tilings. For $D \le 3$ the entropy vanishes.  No
data is available for $D=5$.  For $D=4$ and $6$, data is available
from computer simulations~\cite{4to3,Strandburg91}.  The $6 \ra 3$
case~\cite{Strandburg91} deals with rhombohedron tilings with
icosahedral symmetry.

To estimate the large $D$ limit, we insert oriented membranes into
$d=3$ tilings, as previously we inserted lines into $d=2$
tilings. Membranes of like orientation are forbidden to cross. In the
large $D$ limit, we can neglect contact of like-oriented membranes and
consider the statistics of a single membrane in generic high symmetry
tilings. Let $M_D(\tau)$ denote the number of such membranes that can
be inserted into tiling $\tau$ and $\bar{M}_D$ be the arithmetic mean
of this number. We find that the number of $D \ra 3$ tilings is
\begin{equation}
B_D=\prod_{D'=3}^{D-1} \bar{M}_{D'}.
\end{equation}
analogously to eq.~(\ref{B_D}).  The entropy per tile is the
logarithm of this number divided by the number of tiles.

We obtain an upper bound on the large $D$ entropy of $d=3$ tilings in
a manner similar to our upper bound of $\log{2}$ for for $d=2$.
Recall $2^D$ is an upper bound on the number of directed paths in
$d=2$ tilings, and is actually realized only on special tilings in
which the degree $Z=4$ at all sites, with two incoming and two
outgoing vertices.  Similarly, the number of directed membranes may be
maximal on a special tiling in which the coordination number and
choice numbers are most homogeneous. This is the simple cubic tiling
in which each vertex is six-fold coordinated. The membrane orientation
should be in the plane perpendicular to the (111) axis. At each vertex
there are 18 possible directed membrane configurations. This leads to
$\bar{M}_D \le 18^{N}$ membranes containing $N$ rhombi, a substantial
overestimate because many of these configurations force unique
membrane structures nearby.

A true upper bound is
\begin{equation}
\bar{M}_D \leq e^{\sigma_{3 \ra 2}N},
\hspace{1cm}
\sigma^{D\ra 3} \leq \sigma^{3 \ra 2}
\end{equation}
because the problem of the directed membrane on the simple cubic
lattice is just the $3 \ra 2$ tiling problem. This value is an upper
bound because it neglects contacts, and because generic $D \ra 3$
tilings will be less homogeneous than the simple cubic lattice and
therefore the entropy lower.

\subsection{$D \ra D-1$ tilings}
\label{codim1}

In this section, we study the entropy of strain-free $D \ra D-1$ tilings
where the $D$ different tiles appear with the same fraction $1/D$ and we
derive rigorous lower and upper bounds:
\begin{equation}
\label{bound3d}
{\log 2 \over D} \leq \sigma^{D \ra D-1} < {3 \sigma^{3 \ra 2} \over D} 
\end{equation}
Note that the two prefactors $\log 2 \simeq 0.693$ and $3 \sigma^{3 \ra 2}
\simeq 0.969$ are satisfactorily close and mutually consistent.

To derive the lower bound, we first notice that the elementary tiles
of figure~\ref{flips}, namely hexagons and rhombic dodecahedrons,
perfectly tile the entire two- and three-dimensional space,
respectively. In fact, it is possible to tile any $d$-dimensional
Euclidean space with the shadow $P$ of a unit cube of dimension
$D=d+1$ projected along the $(1,1,\ldots,1)$ direction. To understand
this, consider all the unit cubes of a $D$-dimensional hypercubic
lattice the centers of which belong to the diagonal hyperplane of
equation $\sum x_i =0$. Their projection along the $(1,1,\ldots,1)$
direction is the desired tiling with the tile $P$. Now,
such a polytope $P$ can be tiled in two different ways with the $D$
tiles of the $D \ra D-1$ tilings under consideration, as shown in
figure~\ref{flips}, and independently of its neighbors. If we have a
total of $N$ tiles, we have $N/D$ such polytopes, which provides
$2^{N/D}$ different tilings. Thus the total number of tilings is
bounded below by $2^{N/D}$, which leads to the lower
bound~(\ref{bound3d}) on $\sigma^{D
\ra D-1}$.

To understand the upper bound, we begin with the $4 \ra 3$ case.  In
3-dimensional tilings, we extend the notion of de Bruijn lines and
define de Bruijn surfaces.  The latter
are connected sets of rhombohedral tiles which share a common
face orientation. In $4 \ra 3$ tilings, there are 4 families $F_i$ of
such surfaces, since there are 4 possible edge orientations
$\eh_i^\parallel$. In $d$-dimensional tilings, there are
$(d-1)$-dimensional surfaces defined similarly. Like in the
two-dimensional case, two adjacent surfaces of a same family cannot
cross but can have contacts.

Like in the previous subsection, such a surface in a $4 \ra 3$ tiling
can be put in one-to-one correspondence with a $3 \ra 2$ tiling, just
as in a $3 \ra 2$ tiling, a de Bruijn line can be seen as a directed
random walk on a square grid like a $2 \ra 1$ tiling. Therefore if we
do not take contacts between surfaces into account, a $4 \ra 3$ tiling
is a collection of independent $3 \ra 2$ tilings, corresponding to all
the de Bruijn surfaces of a given family. Of course, this approach
will only provide an upper bound since contacts constraint the
surfaces and reduce the entropy. Now the entropy of $3 \ra 2$ tilings
is known, and a fraction $3/4$ of the tiles of the original $4
\ra 3$ tilings belong to the de Bruijn surfaces of the family
under consideration. Therefore
\begin{equation}
\sigma^{4 \ra 3} < {3 \over 4} \sigma^{3 \ra 2}=0.242.
\label{upper_bound43}
\end{equation}
Now, we can iterate the same argument in any dimension, thus getting that
\begin{equation}
\sigma^{D \ra D-1} < {D-1 \over D} \sigma^{D-1 \ra D-2}
<{D-2 \over D} \sigma^{D-2 \ra D-3}<\ldots<{3 \over D} \sigma^{3 \ra 2} ,
\end{equation}
which leads to the upper bound eq.~(\ref{bound3d}). The
actual $4 \ra 3$ entropy is strictly smaller than our bound
(\ref{upper_bound43}), and its knowledge provides better
upper bounds for larger $D$ tilings. The best numerical estimate of
this entropy is $\sigma^{4 \ra 3} = 0.214 \pm 0.002$ [\cite{4to3}],
which leads to $\sigma^{D \ra D-1} < 4 \sigma^{4 \ra 3} /D = (0.856
\pm 0.008)/D$ for $D \geq 4$, a noticeably better upper bound
than~(\ref{bound3d}).  Similarly, the exact knowledge of any $D \ra
D-1$ entropy would provide a better upper bound for the entropy of
tilings of dimension larger than $D$.

\section{Conclusion}
\setcounter{equation}{0}

This paper studied random tiling models in the limit of high
rotational symmetry. We developed a mean-field theory for tilings
based on the iterative construction of rhombus tilings introduced by
de Bruijn~\cite{debruijn}.  Our main goal is to improve on
the knowledge of the tiling entropy.  Table~\ref{estimate}
summarizes the progress to-date.  Evidently the mean-field
theory is a considerable improvement upon earlier bounds.

The relevant quantity in our mean-field theory is the number of
choices at each vertex, which is related to its coordination number
(degree).  In contrast to prior work that focused on the {\em mean}
number of choices, the current study yields the {\em distribution} of
the number of choices.  Table~\ref{results} demonstrates excellent
agreement with numerical simulation.

We argue that high symmetry tilings become insensitive to phason
strain, unlike finite symmetry tilings.  In other words, phason
elasticity vanishes at large codimension.  In addition to the entropy
and the phason elasticity, we considered finite size corrections and
correlation functions.

Our mean-field theory does not take into account spatial correlations
which should modify slightly the entropy.  For example, if a vertex has
a low degree, it has less chance to be visited in the iterative
process, which modifies the fixed point (\ref{distrib:Nc}) and the
entropy.  Large degree vertices are more likely to be visited and to be
split into two vertices of lower degree.  We expect them to be less
numerous than in our calculation (see table~\ref{results}), which
accounts for the actual lower value of the entropy.

Our mean-field theory could be enhanced by inclusion of
correlations.  Instead of single vertices (with their arms and legs),
one might instead study the statistics of pairs of vertices or even
larger tiling patches and determine how these statistics evolve {\em
via} the iterative process, when they are split into two pieces by a
path going through them.  It would also be useful to predict how the
spatial correlations calculated in the paper perturb the mean-field
entropy.

A second paper~\cite{paperII} follows this one and presents the numerical
Monte Carlo study of large codimension tilings. In particular, it
is demonstrated in this paper that in such tilings, the thermodynamic
limit is restored: fixed boundaries do not result in the lowering
of the entropy like in finite codimension cases. 

\section*{Acknowledgments}
We thank Chris Henley and Pavel Kalugin for useful discussions.
This research is supported in part by the National Science Foundation
under grants DMR-0111198 and INT-9603372 and by the CNRS.

\appendix

\section{Construction of the height function}
\label{hyper}

This appendix constructs~\cite{lpw} the parallel and perpendicular spaces
($\EC^\parallel=\R^2$ and $\EC^\perp=\R^{D-2}$) for $D\ra 2$ tilings
as well as the mapping $h: \EC^\parallel \ra \EC^\perp$. We embed
these spaces in the $D$-dimensional hyperspace $\R^D$. Let the unit
vectors $\{\eh_i\}$ be an orthonormal basis for $\R^D$. Their
normalized projections into $\EC^\parallel$ and $\EC^\perp$ will be
denoted, respectively, by $\eh^\parallel_i$ and $\eh^\perp_i$. Our
first task is to represent these projections in the hyperspace basis.

We define the two-dimensional ``parallel'' space $\EC^\parallel$ by its
orthonormal basis 
\begin{eqnarray}
\label{para-basis}
\hat{x}^\parallel = \sqrt{{2}\over{D}} 
\sum_{j=0}^{D-1} \eh_j \cos{({{\pi j}\over{D}})} \\
\hat{y}^\parallel = \sqrt{{2}\over{D}} 
\sum_{j=0}^{D-1} \eh_j \sin{({{\pi j}\over{D}})}. \nonumber
\end{eqnarray}
Given the basis for $\EC$ we define the projection operator
$P_\parallel$ by
$P_\parallel(u) \equiv (\hat{x}^\parallel \cdot u) \hat{x}^\parallel  +
(\hat{y}^\parallel \cdot u) \hat{y}^\parallel $ and its complement
$P_\perp=1-P_\parallel$. The parallel space vectors $\eh^\parallel_i$
are the projections of suitably scaled hyperspace basis vectors,
\begin{equation}
\label{epara}
\eh^\parallel_i \equiv P_\parallel (s_\parallel \eh_i).
\end{equation}
Note that they form a regular fan
\begin{equation}
\label{fan}
\eh^\parallel_i =
\hat{x}^\parallel \cos{({{\pi i}\over{D}})} +
\hat{y}^\parallel \sin{({{\pi i}\over{D}})}.
\end{equation}
By a ``regular fan'' we mean a collection of
equally spaced vectors arranged in the upper half plane. The union of
these vectors and their negatives forms a $2D$-fold symmetric
star. We set the scale factor $s_\parallel=\sqrt{D/2}$ to normalize
$\eh^\parallel_i$ in eq.~(\ref{fan}).

The $(D-2)$-dimensional ``perpendicular'' space $\EC^\perp$ is the
complement in $\R^D$ of $\EC^\parallel$. Consider the projection of
the scaled hyperspace vector
$s_\perp \eh_i$ into $\EC^\perp$, which we can express as 
\begin{equation}
\label{eperp}
\eh^\perp_i \equiv P_\perp (s_\perp \eh_i) =
s_\perp(\eh_i - P_\parallel \eh_i).
\end{equation}
Its magnitude can be found from
\begin{equation}
\label{perp-norm}
\| P_\perp s_\perp \eh_i \|^2 = s_\perp^2 (1+ \| P_\parallel \eh_i \|^2
-2 \eh_i \cdot P_\parallel \eh_i) \equiv 1.
\end{equation}
We evaluate $\|P_\parallel \eh_i \|^2 = 1/s_\parallel^2 = 2/D$ from
eq.~(\ref{epara}), and we evaluate $\eh_i \cdot P_\parallel \eh_i
= \sqrt{2/D} \eh_i \cdot \eh^\parallel_i = 2/D$ from eq.~(\ref{fan})
together with eq.~(\ref{para-basis}). Then eq.~(\ref{perp-norm}) yields
$s_\perp=1/\sqrt{1-2/D}$. The following result will also be useful:
if $i \neq j$ then
\begin{equation}
\label{double}
\eh_i^\perp \cdot \eh_j^\perp = - {2 \over D-2} \eh_i^\parallel 
\cdot \eh_j^\parallel.
\end{equation}
Indeed, $\eh_i \cdot \eh_j = \delta_{ij} = 
\eh^\parallel_i \cdot \eh^\parallel_j / s_\parallel^2 +
\eh^\perp_i \cdot \eh^\perp_j / s_\perp^2$.

Every vertex $v$ in a rhombus tiling occupies a position of the form
\begin{equation}
{\bf r_\parallel}(v) = \sum_{i=1}^{D} {\cal R}_i(v) \eh^\parallel_i \in
\EC^\parallel
\end{equation}
where the coefficients ${\cal R}_i$ are integers. To find the values
of ${\cal R}_i$ for vertex $v$, choose any vertex as the origin for
$\EC^\parallel$, then follow a path of rhombus edges from the origin
to $v$. Adding to (or subtracting from) ${\cal R}_i$ for each edge
$\eh^\parallel_i$ along the path yields the values of ${\cal R}_i(v)$.
The perpendicular space position of that vertex is defined as
\begin{equation}
h({\bf r_\parallel}) = h(v) = \sum_{i=1}^{D} {\cal R}_i(v) \eh^\perp_i
\in \EC^\perp.
\end{equation}
Values of $h({\bf r})$ for ${\bf r}$ in the interior of a tile can be
defined using linear interpolation on the values just defined for tile
vertices.

The equation of a flat (strain-free) tiling, or of a locally flat
tiling after coarse-graining is defined by $D$ vectors $\vect{m}_i \in
\EC^\parallel$ such that
\begin{equation}
\RC_i(v) = \vect{m}_i \cdot {\bf r}_\parallel,
\end{equation}
the above equality meaning that $\RC_i(v)$ is the closest integer to
the right-hand side quantity. Then the $(D-2) \times 2$ phason
gradient tensor $\vect{E}=\nabla_\parallel h({\bf r}_\parallel)$ is
\begin{equation}
\vect{E}=\sum_{i=1}^D \eh_i^\perp \otimes \vect{m}_i ,
\label{E_and_mi}
\end{equation}
where $\eh_i^\perp \otimes \vect{m}_i$ is a the tensor product of the 
{\em column} vector $\eh_i^\perp$ and the {\em line} vector $\vect{m}_i$.

\end{document}